\documentclass[12pt,a4]{elsart}
\usepackage{natbib}
\usepackage{epsf}
\def\APA{{\tt APACIC++} }
\def\AME{{\tt AMEGIC++} }
\def\AA{{\tt APACIC++/AMEGIC++}}

\newcommand{\nnb}{\nonumber}
\newcommand{\be}{\begin{eqnarray}}
\newcommand{\ee}{\end{eqnarray}}
\newcommand{\bea}{\begin{eqnarray}}
\newcommand{\eea}{\end{eqnarray}}
\newcommand{\barl}{\begin{array}{rl}}
\newcommand{\ba}{\begin{array}}
\newcommand{\ea}{\end{array}}

\newcommand {\vp}{$\vphantom{|^|}$}

\begin{document}
\begin{frontmatter}
\title{{\APA} 1.0\\
{\bf A PA}rton {\bf C}ascade {\bf I}n {\bf C++}}
\author[TU,MPI]{R. Kuhn\thanksref{mail1},}
\author[TECH]{F. Krauss\thanksref{mail2},}
\author[ERIC]{B. Iv{\' a}nyi,}
\author[TU]{G. Soff}
\address[TU]{Institut f\"ur Theoretische Physik, TU Dresden, 
             01062 Dresden, Germany}
\address[MPI]{Max Planck Institut f{\"u}r Physik komplexer Systeme,
              01187  Dresden, Germany}
\address[TECH]{Department of Physics, Technion,
              Haifa 32000, Israel}
\address[ERIC]{Ericsson Telecommunication Ltd.,	
	       H-1037 Budapest, Laborc u.1., Hungary}
\thanks[mail1]{E-mail: kuhn@theory.phy.tu-dresden.de}
\thanks[mail2]{E-mail: krauss@physics.technion.ac.il}
\begin{center}
{\it\small (submitted to Computer Physics Communications)}
\end{center}

\begin{abstract}
\APA is a Monte--Carlo event--generator dedicated for the simulation of electron--positron annihilations into jets.
Within the framework of \APA, the emergence of jets is identified  with the perturbative production of partons as governed 
by corresponding matrix elements. In addition to the build--in matrix elements describing the production of two and 
three jets, further programs can be linked allowing for the simultaneous treatment of higher numbers of jets. \APA hosts 
a new approach for the combination of arbitrary matrix elements for the production of jets with the parton  shower, 
which in turn models the evolution of these jets. For the evolution, different ordering schemes are available, namely 
ordering by virtualities or by angles. At the present state, the subsequent hadronization of the partons is accomplished by 
means of the Lund--string model as provided within {\tt Pythia}. An appropriate interface is provieded.

The program takes full advantage of the object--oriented features provided by C++ allowing for an equally 
abstract and transparent programming style. 
\end{abstract}
\end{frontmatter}

\newpage
\section*{Program Summary}
{\it Title of the program :} \APA, version 1.0\\

{\it Program obtainable from :} CPC Program Library and upon request, homepage is under construction\\

{\it Licensing provisions :} none\\

{\it Operating system under which the program has been tested :} UNIX, LINUX, VMS\\

{\it Programming language :} C++, some interfaces in Fortran77\\

{\it Separate documentation available :} in preparation\\

{\it Keywords :} QCD, standard model, gauge bosons, Higgs physics, $e^+e^-$ annihilations, 
                          jet production, parton shower\\

{\it Nature of the physical problem: } With rising energies, the final state in high--energy
                        electron positron--annihilations becomes increasingly complex. The number of jets
                        as well as the number of observable particles, leptons, hadrons and photons, 
                        increases drastically and prevents any analytical prediction of the full final state.
                        In addition, the transformation of the partons of perturbative quantum field heory into the 
                        experimentally observable hadrons is so far not understood on a quantitative level.
                        Both obstacles prevent any attempt to bring the underlying theory in direct contact
                        with the final states by analytical methods.\\

{\it Method of solution: } \APA produces complete $e^+e^-$--events on a level suitable for direct 
                        comparison with experiment. The events are generated using Monte--Carlo methods 
                        and by dividing their simulation into well--separated steps. \APA concentrates in its
                        event generation on the hard subprocess producing jets and the subsequent
                        parton shower describing their evolution. For the production of jets, interfaces to various matrix element 
                        generators are provided. The fragmentation into hadrons and their subsequent decays 
                        are left for well--defined models encoded in already existing Fortran programs. 
                        Suitable interfaces are supplemented.
\newpage
\tableofcontents
\newpage

\section{Introduction}

During the last decades, the investigation of $e^+e^-$--collisions with ever rising energies
provided one of the central laboratory frames of particle phenomenology. Confronting
experimental results and theoretical predictions led to a large number of conclusions
covering a good part of what is known nowadays as the Standard Model. Without
going into great detail, these results include

\begin{enumerate}
\item{establishing QCD as the best model underlying strong interactions by 
          \begin{enumerate}
          \item the discovery of the gluon in three--jet events \cite{Petra}, 
          \item the measurement of the Casimir operators $C_F$ and $C_A$ \cite{Opal}
                   of the fundamental and adjoint representation of the group $SU(3)$ defining
                   the gauge sector of QCD as well as the determination of the normalization of their generators, and
          \item the confirmation of the correct running of $\alpha_s$ in a large interval of scales \cite{alphaS}.
          \end{enumerate} }
\item{highly precise measurements within the electroweak sector of the
          Strandard Model, for example masses and widths of the gauge bosons \cite{WZmasses}, thus
          \begin{enumerate}
          \item establishing the Standard Model as an extremely reliable model even at quantum level, 
                   at least at the scales under investigation,
          \item put increasingly severe bounds on the mass of the so far unobserved
                    Higgs--boson \cite{Higgs}, and, last but not least
          \item constraining considerably the parameter space and the models for physics
                    beyond the Standard Model.
          \end{enumerate}} 
\end{enumerate}

Unfortunately, the mutual mapping of theoretical predictions and experimental results onto each other
prove far from being trivial. Three rather different reasons give rise to these difficulties, namely :

\begin{enumerate}
\item{Quite a large number of $e^+e^-$--annihilation processes involving the full c.~m. energy of the
          colliding beam particles end up with strong interacting final states. The confinement property
          of QCD \cite{Confin} then enforces the transition from the partons, the particles of perturbation theory, quarks and gluons,  
          to the observable hadrons detected in the experiments. At best this transition is understood merely 
          qualitatively, and it is fair enough to claim, that so far there is no quantitative model starting from first 
          principles, i.~e. derived from the Lagrangian of QCD. Instead, currently the only approach is to describe 
          fragmentation with purely phenomenological models with essentially free parameters to be tuned
          to existing experimental data.}
\item{On the other hand, even at the parton level, events usually accommodate a prohibitive large number of
          particles to be dealt with analytically. Consequently, the standard methods of perturbative field theory,
          i.~e. summing all Feynman--amplitudes, fail badly in any attempt to describe the partonic
          ensemble before the fragmentation regime is entered. The only viable way out of this dilemma so far 
          is to abandon this method of calculations yielding an exact result in the full phase space. Instead,
          one concentrates on the dominant regions of soft and collinear particle production common
          to field theories with -- nearly -- massless particles. Expanding around the appropriate limits, the 
          production processes factorize neatly into single binary particle decays, which can be resummed.
          Additionally, this approach provides some insight into the space--time structure of strong interactions. 
          Moreover, the parton shower picture leads itself to an implementation in terms of a computer program
          using some Monte Carlo approach.}
\item{This approach, however, in most cases is not capable of describing the bulk of interesting signatures
          involving more than two or three particles produced in a hard subprocess. This is due to the
          fact, that any expansion around soft and collinear limits fails by construction when attemting to
          describe multijet events with high--energy particles and large opening angles. In fact, for such 
          processes, the only possibility yielding exact results are the corresponding matrix elements.  In principle
          they can be evaluated with systematically increasing accuray when going to higher orders of
          perturbation theory. In practice, in most cases results exist only at the quantum level, i.~e. at the
          one--loop order.

          Obviously, this unpleasant situation when trying to describe multijet production
          needs to be resolved.}
\end{enumerate} 

As already mentioned, a popular and fruitful approach to handle the difficulties encountered above is the
use of computer programs, so called Monte Carlo event generators, to simulate full events. The basic
strategy of such programs can be headlined as {\em divide et impera}. In other words, usually such programs
divide individual events into single, disjunct stages and treat them separately. In doing so, the algorithms might miss
possible non--trivial correlations between different steps, like the interference of photon radiation 
off the initial and final state. On the other hand, apart from being the only working approach so far, this strategy 
allows for independent tests of each step by comparing with suitable sets of data. 

In this paper a new event generator, \APA, is presented which currently is capable to simulate the essentials of 
$e^+e^-$--events at LEP--II energies and beyond. Two features of \APA mark the important differences
compared to other popular codes like {\tt Ariadne}\cite{Ariadne}, {\tt
Herwig}\cite{Herwig}  and {\tt Pythia}\cite{Pythia} :

\begin{enumerate}
\item \APA is written from scratch in the modern language {\tt C++} \cite{cplus}. Its objec--\-oriented features allow for an abstract 
         and comprehensible programming style and an increased control of the data flow within the program.
         We want to express our strong opinion, that this results in an user--friendly code.
\item Within \APA, a generically new approach to combine arbitrary matrix elements and parton showers has been 
         formulated and implemented \cite{KKS99}. Together with the new matrix
         element generator \AME\cite{AMEGIC} , this will enable \APA to
         simulate most features of current and future $e^+e^-$--experiments.
\end{enumerate} 

So, the outline is as follows. In the next section, Sec.\,\ref{PHYS} we briefly introduce the major physical concepts 
encoded within \APA. We put some emphasize on new algorithms only, like for instance the procedure for combining matrix 
elements and the parton shower. In Sec.\,\ref{PROG} we outline in some detail the class structure of \APA. There, we feel
justified to go into some detail for the benefit of those readers not too familiar with {\tt C++}. The following part, Sec.\,\ref{inst} is devoted
to the implementation of \APA and provides a rather concise description of the prerequisites and steps eventual users
have to follow. Additionally, some of the parameters and switches steering \APA are described. While Sec.\,\ref{Summ}
summarizes with some final remarks including our aims for \APA in the future, at the end of the paper we have provided an
exemplatory test run output.

\newpage
\section{Physics Overview\label{PHYS}}

\begin{figure}[h]
\centerline{\epsfxsize=8cm\epsffile{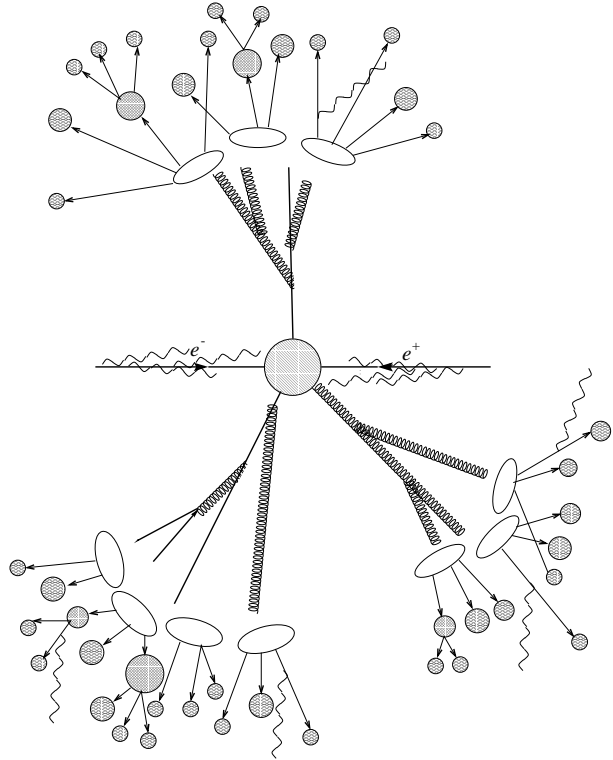}}
\caption{\label{stagfig} Scetch of an $e^+e^-$--annihilation into jets.
The wiggly lines represent the photons of the initial state radiation,
the thick shaded blob stands for the hard subprocess, here the production
of three jets. The secondary parton radiation accounts for the inner--jet evolution,
whereas the fragmentation is indicated by the ellipses with further hadronic decays indicated.}
\end{figure}

In this section, we would like to summarize the physics encoded in the new
event generator \APA. In its present state, \APA is capable to describe $e^+e^-$
initiated processes at LEP energies and beyond putting a strong emphasis on strong interacting 
final states. Such processes, e.~g. $e^+e^-\to$ jets, can be 
modelled in terms of the following steps, see Fig.\,\ref{stagfig} for comparison :
\begin{enumerate}
\item{Initially, two beam particles, i.~e. the electron positron pair, are approaching
each other, usually head--on--head. Eventually they radiate photons, which are 
predominantly soft and collinear. Thus, as a first approximation, this initial state 
radiation of photons off the electrons merely changes the energies, but not the direction 
of the beam particles.}
\item{With a c.~m.--energy, which is reduced accordingly, the electron positron pair
interacts producing varying numbers of primary partons. The main properties of this hard 
subprocess and the kinematical distribution of the primary partons determine the overall 
features of the event. Therefore it is reasonable to concentrate in this step on final state 
particles with comparably high energies and large relative angles, i.~e. jets.}
\item{The jets produced in the hard subprocess experience an evolution from the 
hard scales of their production down to the relatively soft scales of hadronization. 
In the progress of their evolution, the partons loose their timelike virtual mass via multiple
splitting into pairs of secondary partons where each of the decay products is also provided 
with -- lower -- virtual masses and might decay further. This parton shower stops at some minimal 
virtual mass $q_0^2$ of the order of a few $\Lambda_{\rm QCD}$.}
\item{The resulting parton ensemble is now fragmented into the observable colour--neutral
hadrons. Since this is an essentially non--perturbative process, there is a definite lack of 
quantitative understanding starting from first principles. Thus, parameter dependent phenomenological
models have to be employed for the description of hadronization. 

However, many of the produced hadrons are unstable and decay further.}
\end{enumerate}
In this context, a comment is in order. As a matter of fact, the parameters of the 
hadronization model employed depend strongly on the energy scale related to the onset
of fragmentation. In this sense, the two basic reasons for the implementation of the parton 
shower in event generation are
\begin{enumerate}
\item{to give a better description of inner--jet features, and}
\item{to provide the hadronization model with an universal energy scale $q_0^2$
for its onset, which is independent of the c.~m.--energy of the process.}
\end{enumerate}
In this sense, the {\em parton shower guarantees the universality of the hadronization model}.

One of the long--standing obstacles of event generation for high--energy processes 
is to combine the matrix elements describing the hard process of jet production to the parton 
shower. In \APA a new algorithm was developped and implemented resolving this problem.

Apparently, the steps outlined above follow some remnant idea of time ordering and, in 
addition, they are characterized by roughly disjunct energy regimes. Note, that for the sake of
compact expressions, here and in the following we denote by partons indiscriminately any 
elementary particle, i.~e. leptons and the electroweak gauge bosons in addition to the quarks and 
gluons.

In the rest of this section, we will discuss the stages of an event named above in a slightly 
rearranged way. Since most of the physics features encoded in \APA are already covered in a very 
detailled manner in various publications and textbooks\cite{ESW96}, we will restrict ourselves to quite
a scetchy presentation of these issues and corresponding references. On the other hand, the 
new approach for the combination of matrix elements and parton showers represents original 
work and therefore more care is spent on the discussion of this part.

\subsection{Matrix elements\label{physXS}}

We start our tour de force through the physics encoded within \APA with a discussion of the
hard underlying process. Here, differences of \APA to other frequently used event generators,
like {\tt Pythia} or {\tt Herwig} become most apparent. Going beyond single exclusive channels, 
these generators usually start with $e^+e^-\to q\bar q$, populating the phase space for particle 
emission with help of the suitably corrected and set-up parton shower, see Subsec.\,\ref{physPS}. 

In contrast, \APA divides the phase space into two disjunct regions \cite{Sey95} by means of the notion of jets
\cite{KKS99a,KKS99b}.
Popular jet measures available within \APA are the Jade-- \cite{Jade}
and the Durham--scheme \cite{Durham}, defining two particles to belong two different jets, if 
\bea
\begin{array}{rcll}
2E_iE_j(1-\cos\theta_{ij}) & > & y_{\rm cut} s_{ee} \;\;\;\;\;\; (\mbox{\rm Jade}) \\
2\mbox{\rm min}\{E_i^2,E_j^2\}(1-\cos\theta_{ij}) & > & 
y_{\rm cut} s_{ee} \;\;\;\;\;\; (\mbox{\rm Durham})\,.
\end{array}
\eea
Within \APA, the user predefines an initial $y_{\rm cut}$, in the following called $y_{\rm ini}$, 
and the corresponding scheme. Then emissions characterized by a $y>y_{\rm ini}$ are
described by means of the corresponding matrix elements squared, thus identifying the outgoing 
partons with jets according to the initial definition. The complementary regime of parton radiation 
with $y<y_{\rm ini}$ is covered by the parton shower. This division of phase space in
two region is maintained in \APA even for varying numbers of jets, i.~e. the simultaneous
generation of events in all channels accessible. Then, the selection of the final state proceeds
in four steps, namely :
\begin{enumerate}
\item{During the initialization of \APA total cross sections for each channel under inspection 
          in dependence on the jet--definition are either read in or calculated.
          To account for the impact of higher order corrections in QCD channels, some scale 
          factors $\kappa_{n_j}$ are introduced to modify each $n_j$--jet cross section $\sigma_{n_j}$
          by replacing the corresponding prefactor $\left[\alpha_s(s_{ee})\right]^{n_j-2}$ 
          with $\left[\alpha_s(\kappa_{n_j} s_{ee})\right]^{n_j-2}$. 
Similar treatments can be found in \cite{Pythia,SPINORS2}.
Within \APA the running of 
          $\alpha_s$ is taken in leading order.}
\item{Now, $n_j$--rates $R(n_j)$ are defined. \APA provides four different schemes, 
          a ``direct'' one, two ``rescaled'' ones and a ``resummed'' one. 
          Defining $\sigma_{\rm had} = \sigma_{ee\to q\bar q}$
          and concentrating on events mediated by one intermediate photon or $Z$--boson,
          the direct one reads
           \bea\label{Rdir}
           R(n_j)^{\rm dir.}  &=&  \frac{\sigma_{n_j}}{\sigma_{\rm had}}\;,\nnb\\
           R(2)^{\rm dir.} &=& 1-\sum\limits_{n_j>2} R(n_j)^{\rm dir.}
           \eea
           and the two rescaled schemes are
           \bea\label{Rres1}
           R(n_j)^{\rm res1} &=& R(n_j)^{\rm dir.} -
                                                    \sum\limits_{n_k>n_j} R(n_k)^{\rm dir.}\;,\nnb\\
           R(2)^{\rm res1} &=& 1-\sum\limits_{n_j>2} R(n_j)^{\rm res1}
           \eea
           and
           \bea\label{Rres2}
           R(n_j)^{\rm res2} &=& R(n_j)^{\rm dir.} 
                                                   \prod\limits_{n_k>n_j}\left(1-R(n_k)^{\rm res2}\right)\;,\nnb\\
           R(2)^{\rm res2} &=& 1-\sum\limits_{n_j>2} R(n_j)^{\rm dir2}
           \eea
           where the first scheme obviously treats $n_j+k$--jet configurations as subsets of 
           $n_j$--configurations and the effect of the scale factors $\kappa_{n_j}$ is already included.  

           In the fourth scheme, the resummed one, the matrix elements squared giving rise to the jetrates
           are combined with jetrates in the so--called NLL-scheme \cite{DURHAM} relying on Sudakov form factors
           \cite{Sud}. These Sudakov form factors have an interpretation as the probability of no observable
           branching between two scales, see Subsec.\,\ref{physPS} and in leading logarithmic order 
           they are given by
           \bea\label{SudNLL}
           \Delta_q^{\rm NLL}(Q_{\rm ini},Q) &=& \exp\left[\int\limits_{Q_{\rm ini}}^Q
                                                                          \,dq\Gamma_q(q,Q) \right]\nnb\\
           \Delta_g^{\rm NLL}(Q_{\rm ini},Q) &=& \exp\left[\int\limits_{Q_{\rm ini}}^Q
                                                                          \,dq\left(\Gamma_g(q,Q) +\Gamma_f(q)\right)\right]\nnb\\
           \Delta_f^{\rm NLL}(Q_{\rm ini},Q) &=& \frac{\left[\Delta_q^{\rm NLL}(Q_{\rm ini},Q)\right]^2}
                                                                                      {\Delta_g^{\rm NLL}(Q_{\rm ini},Q)}
           \eea
           with the NLL--splitting functions $\Gamma$ representing in the same approximation the
           branching probabilities for $q\to qg$, $g\to gg$ and $g\to q\bar q$, respectively,
           \bea\label{splitNLL}
           \Gamma_q(q,Q) &=& \frac{2C_F}{\pi}\frac{\alpha_s(q)}{q}
                                               \left(\log\frac{Q}{q}-\frac34\right)\nnb\\
           \Gamma_g(q,Q) &=& \frac{2C_A}{\pi}\frac{\alpha_s(q)}{q}
                                               \left(\log\frac{Q}{q}-\frac{11}{12}\right)\nnb\\
           \Gamma_f(q,Q) &=& \frac{n_f}{3\pi}\frac{\alpha_s(q)}{q}
           \eea
           Then, for example, the two--and three--jetrates read in NLL--approximation
           with $Q_{\rm ini}^2 = y_{\rm ini}s_{ee}$ and $Q^2 = s_{ee}$ 
           \bea\label{RNLL}
           R_2(Q_{\rm ini},Q) &=& \left[\Delta_q(Q_{\rm ini},Q)\right]^2\nnb\\
           R_3(Q_{\rm ini},Q) &=& 2\left[\Delta_q(Q_{\rm ini},Q)\right]^2
                  \int\limits_{Q_{\rm ini}}^Q\,dq\Gamma_q(q,Q)\Delta_g(Q_{\rm ini},q)
           \eea
           and they are combined with the direct rates above by expanding in $\alpha_s$ and replace 
           the coefficients for the corrsponding powers of $\alpha$ with the direct rate above, 
           Eq.\,(\ref{Rdir}). 
           Note that in this scheme, the various scale factors $\kappa_s$ are forced to be equal to 1.
          
           The jetrates defined above in the three schemes hold true for pure QCD final states and one
           intermediate photon or $Z$--boson. Including more electroweak gauge bosons with decays
           resulting in at least four fermions, the situation changes. Then two subsets are defined, 
           where all channels with at least four fermions in the final state are excluded from the 
           QCD--subset and added to the electroweak set EW. The cross section for this last set is given 
           by the sum of all contributing channels, the cross section for the QCD set is still assumed to 
           be $\sigma_{\rm had}$.}
\item{During the initializtion of the individual events, first the subset, either QCD or EW, is 
          chosen according to the cross section. In case of QCD the number of jets is then determined
          via the corresponding jetrate, $R(n_j)$ given above.}
\item{Having defined the number of jets of the event or its membership to the EW--set, the flavour 
          constellation is picked according to the relative weight of its contribution to the jetrate
          or the electroweak subset.} 
\end{enumerate}
\APA itself provides expressions for two processes only, namely
\bea
e^+e^-\to q\bar q\;\;\mbox{\rm and}\;\;\; 
e^+e^-\to q\bar qg\,,
\eea
where both photon-- and $Z$--exchange and quark masses can be taken
into account \cite{Dok94}.
In addition, \APA includes interfaces to a number of matrix element generators allowing for
a considerably larger class of processes, see Table\,\ref{MEav.}. For further details on those
generators we refer to the corresponding literature.
\begin{table}[h]
\begin{center}
\begin{tabular}{|l|c|c|c|l|}
\hline
Generator                                   & $\%$ of jets & LO/NLO & Quark masses & Comments \\\hline
\AME\cite{AMEGIC}                  & $\le 5$          &    LO       & yes      & preferred choice, full SM\\
{\tt Debrecen}\cite{Debrecen} & $\le 4$           & NLO       & no       &  QCD only \\
                                                    & $\le 5$           &   LO       & no       &                     \\
{\tt Excalibur} \cite{Excalibur}  & $=4$               &   LO        & no      &  full SM, no Higgs.\\
\hline
\end{tabular}
\caption{\label{MEav.} Matrix element generators to be interfaced with \APA.}
\end{center}
\end{table}

\subsection{Initial state radiation\label{physISR}}

Running \APA with \AME or the built--in matrix elements, 
there is an option to include the effect of initial state radiation of
photons off the electrons. Presently, both programs allow only for quite a simple approximation
in the description of this effect, namely the structure function approach \cite{ISR}. In this approach,
the photons are emitted on--shell strictly collinear, i.~e. parallel to the beam axis and thus, they
merely reduce the energy of he incoming electron. With $x$ the energy of the electron in units 
of its beam energy and $m_e$ the electron mass, the structure function has the form
\bea
\Phi(x) &=& \frac{\exp\left(-\beta_E\gamma_E+\frac{3}{8}\beta_S\right)}
                            {\Gamma\left(1+\frac12\beta_{\rm exp}\right)}
                     \beta(1-x)^{\beta_{\rm exp}/2-1} - \frac14\beta_H(1+x)  + O(\alpha^2)\,,
\eea
where $\beta_{\rm exp} = \beta$ and the two other $\beta_i$ encountered are either 
$\beta$ or $\eta$,
\bea
\beta &=& \frac{2\alpha}{\pi}(L-1)\;,\;\; \eta=\frac{2\alpha}{L}
\;,\;\; L = \log\frac{s_{ee}}{m_e^2}\,,
\eea
each choice representing a different parametrization. Within \APA and \AME, this structure 
function $\Phi(x)$ is encoded up to third order in $\alpha$, the default setting is the 
so--called $\beta$--choice,
\bea
\beta_{\rm exp} = \beta_H = \beta_S = \beta
\eea
and $\Phi$ up to $\alpha^2$. 

Including the effect of initial state radiation in this framework merely adds two more variables
to the phase space integral to be performed, namely the energy fraction $x_{1,2}$ of the electron 
and the positron, respectively, but it does not alter the way, the jet constellation of the events
is determined.

\subsection{Combining matrix elements and parton showers\label{physXSPS}}

Following \APA in the process of event generation, we turn now to the issue of combining the 
matrix elements described above, see Subsec.\,\ref{physXS}, with the subsequent parton shower to be 
adressed later in \ref{physPS}. Assuming LO matrix elements for jet production only, the new algorithm 
covering this task proceeds in the following steps \cite{KKS99}:
\begin{enumerate}
\item{Having chosen the number of jets and the flavour constellation in the fashion already
          described above, the kinematical constellation is determined according to the corresponding
          matrix element with a hit--or--miss method. For this purpose one needs some maximal
          value limiting it from above. This maximum has already been found during the Monte--Carlo 
          evaluation of the cross sections, which sampled the matrix element over the available 
          phase space, and it has been stored.}
\item{\APA provides three different schemes off additional weights multiplying the numerator of the
         hit--or--miss method. They are introduced to model some of the higher order effects on top 
         of the LO matrix element. The first scheme is a direct one, which does not alter at all the 
         distributions given by the matrix element, the second one includes the effect of running 
         $\alpha_s$
         \bea
         w^{\rm dir.}(n_j) &=& 1\;,\\
         w^{\alpha_s}(n_j) &=& \left[\frac{\alpha_s(y_{\rm min} s_{ee})}
                                                               {\alpha_s(y_{\rm ini} s_{ee})}\right]^{n_j-2}
         \eea
         with $y_{\rm min} = \mbox{\rm min}_{i,j}\{ y_{ij} \}$, the minimum of all values $y$ between two
         jets $i$ and $j$ of the event and $y_{\rm ini}$ the $y$ used for the initial jet definition. 

         The third scheme is the most involved one and employs additionally Sudakov form factors in 
         the NLL--approximation, see Subsec.\,\ref{physXS}. Their interplay depends in a non--trivial 
         way on the event structure and the resulting weight again resums in NLL--approximation the 
         effect of multiple soft and collinear emissions of secondary partons. 

         More specifically, this last weight is constructed recursively. Starting from a $n_j$--jet 
         configuration with $n_j$ four momenta, the two momenta $i$ and $j$ with the smallest $y_{ij}$ 
         are clustered yielding a new four momentum related to some internal line. The clustering is 
         repeated until only two internal quark lines remain. Then, each internal line is weighted with a 
         ratio of Sudakov form factors, representing the probability that no emission resolvable at the 
         scale associated with the initial jet definition, $y_{\rm ini}$, takes place between the upper and 
         lower scales of the line, which are defined via the corresponding values $y^{u,l}$. Outgoing
         lines in contrast yield merely a single Sudakov form factor with the upper scale given by the
         $y^u$ of their production and the lower scale $y_{\rm ini}$. As an illustrative example, consider the three jet 
         configuration displayed in Fig.\,\ref{exthree}
         \begin{figure}
         \centerline{\epsfxsize=4.cm\epsffile{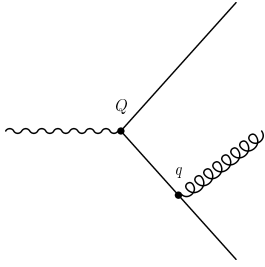}}
         \caption{\label{exthree} Typical three--jet configuration.}
         \end{figure}
         with the corresponding ``resummed'' weight
         \bea
         w^{\rm res} = \frac{\alpha_s(q^2)}{\alpha_s(Q_{\rm ini}^2)}
         \Delta_q(Q_{\rm ini}, Q) \frac{\Delta_q(Q_{\rm ini}, Q)}{\Delta_q(Q_{\rm ini}, q)}
           \Delta_q(Q_{\rm ini}, q) \Delta_g(Q_{\rm ini},q)\,,
         \eea
         where
         \bea
         Q= \sqrt{s_{ee}}\;,\;\;
         q = \sqrt{\mbox{\rm min}\{y_{qg},y_{\bar qg}\} s_{ee}}\;,\;\;
         Q_{\rm ini} = \sqrt{y_{\rm ini} s_{ee}}\,.
         \eea
          For more details on this scheme, we refer the reader to \cite{CKKW00}. There a proof is also 
          given, that when initializing physically meaningful jetrates at $y_{\rm ini}$ this algorithm 
          reproduces the jetrates at arbitrary larger values of the resolution parameter $y_{\rm cut}$ in 
          leading logarithmic approximation.
          }  
\item{Having determined the proper kinematical configuration in one of the three schemes introduced 
          above, the colour constellation of the event is chosen. This is accomplished by defining 
          relative probabilites for each parton history representing a specific colour flow. \APA provides 
          four schemes, the first two employing -- if available -- the Feynman--amplitudes related to 
          the diagrams. Here, up to some appropriate normalization, the relative probabilities $P_i$ for 
          each specific colour history $i$ related to some colour flow as given in a diagram/amplitude 
          $M_i$ reads
          \bea\label{dirprop}
          P_i^{\rm dir.1} \sim \left| M_i^2 \right|\;\;\mbox{\rm or}\;\;
          P_i^{\rm dir.2} \sim \left| M_i \sum\limits_j M_j^* \right|\;,
          \eea
          respectively. 

          The third scheme employs the language of the parton shower in a fashion similar to the one
          presented in \cite{AS98}. Here, all possible
          ways to reach the given configuration via a chain of $1\to 2$--branchings is constructed 
          recursively. Each internal line contributes a factor $1/t$, where $t$ is the invariant mass
          of the line, and each splitting $a\to bc$ is represented by the corresponding splitting
          function $P_{a\to bc}(z)$. Note, that since all four momenta of the final state are known, the
          kinematical parameters $t$ and $z$ can easily be determined. As an illustrative example, 
          consider the four--jet configuration depicted in Fig.\,\ref{exfour}.
          The relative probability in this scheme reads
           \bea\label{partprop}
           P &=& \frac{1}{t_1} P_{q\to gq}(z_{1\to 34}) \frac{1}{t_3} P_{g\to gg}(z_{3\to 56})\;,\nnb\\
           t_1 &=& p_1^2 = (p_4+p_5+p_6)^2\;,\;\;
           z_{1\to 34} = \frac{E_3}{E_1}\;,\nnb\\       
           t_4 &=& p_4^2 = (p_5+p_6)^2\;,\;\;
           z_{4\to 56} = \frac{E_5}{E_4}\,.
           \eea
          \begin{figure}[h]
          \centerline{\epsfxsize=4.cm\epsffile{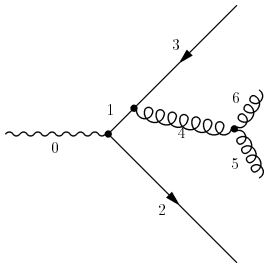}}
          \caption{\label{exfour} Typical four--jet configuration.}
          \end{figure} 
           The fourth scheme applies only, if the kinematical configuration has been chosen in the
           resummed algorithm including the Sudakov form factors, described above. Then the colour
           configuration is determined as the one yielding the most advantageous clustering.
          }
\item{The final step is to provide timelike virtual masses to the outgoing partons, which so far
          have been on their mass shell. This is accomplished with the regular parton shower 
          algorithm described below. The corresponding upper scales for each parton are then given by
          the virtual mass related to the splitting before, i.~e. $t_4$ for partons $5$ and $6$,
          $t_1$ for parton $3$ and $Q^2$ for parton $2$ in the exemplary graph above. Since the 
          subsequent parton shower is limited to model the inner--jet evolution, in the determination of 
          the lower scale a veto is applied on unwanted virtual masses producing an additional jet, i.~e. 
          on virtual masses which translate in scales larger than $y_{\rm ini}$.   

          To guarantee local four momentum conservation when providing virtual masses,
          the corresponding four momenta are slightly reshuffled. In close analogy to the algorithm            
          within the parton shower, the new momenta $p_{b,c}^{\rm cor.}$ in terms of the
          original ones $p_{b,c}^{(0)}$ read
          \bea\label{Kincor}
         p_{b,c}^{\rm cor.} = p_{b,c}^{(0)} \pm\left(r_cp_c^{(0)}-r_bp_b^{(0)}\right)\,,
         \eea
         where the offsprings $b$ and $c$ stem from the internal line $a$. The factors 
         $r_{b,c}$ are then given by
         \begin{itemize}
         \item{Case 1: b is an internal line, c is outgoing.
                   \bea\label{Kincor10}
                   r_b &=& \frac{t_a+(t_c-t_b)-\lambda}{2t_a}\nnb\\
                   r_c &=& \frac{t_b(t_b-t_c+\lambda)-t_a(t_a-t_c-\lambda)}
                   {2t_a(t_b-t_a)}
                   \eea}
          \item{Case 2: b and c are outgoing.
                    \bea\label{Kincor11}
                    r_{b,c} &=& \frac{t_a \pm(t_c-t_b)-\lambda}{2t_a} 
                    \eea}
          \end{itemize}
          $\lambda$ is
          \bea
          \lambda = \sqrt{(t_a-t_b-t_c)^2-4t_bt_c}\,.
          \eea
         }
\end{enumerate}
This closes the presentation of the new algorithm to combine matrix elements and parton showers
as provided in \APA and we turn our attention to the subsequent parton shower modelling the 
inner--jet final state radiation. 

\subsection{Final state radiation : The parton shower\label{physPS}}

The common approach to model the pattern of multiple emissions of partons constituting 
the final state radiation is the parton shower picture \cite{ESW96}. Basically, it involves the 
concentration on the soft and collinear regime of phase space housing the largest contributions and 
thus the bulk of emissions. Expanding each individual parton splitting around the
corresponding soft and collinear limits results in a factorization of the full -- presumably 
complicated -- radiation structure into a chain of independent decays, which can be treated in a
probabilistic manner. In this framework, the leading logarithms are resummed in two 
different schemes employing different order parameters, namely the ordering by virtual masses
in the leading log--scheme (LLA), which is inspired by the well--known DGLAP--equation
\cite{DGLAP} , and the ordering by angles in the modified leading log--scheme (MLLA)
\cite{Dok91}. The important effect of coherence \cite{Dok91,Chud} in the parton shower is provided 
in the first scheme by an appropriate veto on rising opening angles in subsequent parton splittings 
\cite{Ben87a,Ben87b}, in the second scheme this effect is incorporated in a natural fashion. 

The parton shower in both schemes is organized by means of Sudakov form factors \cite{Sud}
\bea\label{Suddef}
\Delta_{a\to bc}(t_0(a),t) \equiv \exp\left[-\int\limits_{t_0(a)}^{t}\,\frac{dt'}{t'}
     \int\limits_{z_1(t')}^{z_2(t')}\,dz\frac{\alpha_s(p_\perp(z,t'))}{2\pi}\,P_{a\to bc}(z) \right]\,,
\eea
where $P_{a\to bc}$ is the splitting function related to the decay $a\to bc$.  Eq.\,(\ref{Suddef}) yields 
the probability, that no resolvable branching $a\to bc$ occurs between the scales 
$t$ and $t_0(a)$, which is usually taken as the infrared cut--off of the parton shower. Consequently, 
ratios $\Delta(t_0,t_1)/\Delta(t_0,t_2)$ are identified as the probability that no branching resolvable at 
the infrared scale $t_0$ happens between $t_1$ and $t_2$. Within event generators such ratios are 
constructed and compared with random numbers and determine the -- decreasing -- sequence of 
scales in accordance with the ordering schemes named above.  

In terms of the Sudakov form factors, LLA and MLLA differ in the interpretation of the scale parameter 
$t'$. In LLA $t'$ is the timelike virtual mass of the decaying parton, whereas in the MLLA,
$t' = E_a^2\theta_{a\to bc}$, the scaled opening angle. Consequently LLA and MLLA differ in the 
definition of the relative transversal momentum identified with the scale of $\alpha_s$ and the 
boundary conditions for $z$. 
\bea
p^2_\perp &\stackrel{\rm LLA}{\Longrightarrow} & z(1-z) t' 
                      \stackrel{\rm MLLA}{\Longrightarrow} z^2(1-z)^2 t'\nnb\\ 
z_{1,2}^{\rm LLA} &=& \frac12\pm \frac12\sqrt{1-\frac{4 t_0(a\to bc)}{t'}}\nnb\\
\sqrt{\frac{q^2_0}{t}} \le & z^{\rm MLLA} & \le 1 - \sqrt{\frac{q^2_0}{t'}}\;.
\eea
In \APA, both ordering schemes for the parton shower are available, the additional 
angular veto in the LLA--scheme can be switched off by the user. Note, that within \APA the first
splitting of a parton within the shower is {\em always} performed in the LLA--scheme. This is due to the
fact, that by construction MLLA is only applicable in the region of small angles, which might
not yet be reached for the first branching.

However, in \APA running with the LLA--scheme each parton leaves the parton shower with a 
flavour dependent virtual mass,
\bea
t_0(f) &=& \mbox{\rm min}\{q_0^2, m_f^2\}\;,
\eea
thus restricting the minimal virtual mass for each specific decay channel via 
\bea
4t_0(a\to bc) &=& \left[\sqrt{t_0(b)}+\sqrt{t_0(c)}\right]^2\,.
\eea
This results in restrictions $4t_0(g\to bb) \ge 4m_b^2$ for gluons splitting into two $b$--quarks
and $4t_0(b\to bg)\ge (m_b+q_0)^2$ for decays $b\to bg$. Therefore, within \APA the Sudakov 
form factors are constructed as the sum of form factors corresponding to the individual possible
decays. In the algorithm of \APA individual splittings proceed as follows 
\begin{enumerate}
\item{Starting with the upper scale $t_1$ first the virtual mass of the next observable decay of 
          parton $a$, $t_2$ is determined via the comparison of a random number $\#R$ with the 
          appropriate sum of Sudakov form factors,
           \bea
           \#R \stackrel{!}{=}  \frac{\sum_{bc} \Delta_{a\to bc}[t_0(a\to bc),t_1]}{\Delta[t_0(a\to bc),t_2]} 
           \Longrightarrow t_2.
           \eea
          }
\item{Then the energy fraction $z'$ is determined according to the sum of splitting functions
          with a hit--or--miss method. Here, first a $z'$ is chosen uniformly in the maximal allowed 
          range of all decay channels,
           \bea
           \mbox{\rm min} \{z_1(a\to bc)\} \le z' \le  \mbox{\rm max} \{z_2(a\to bc)\}   
           \eea
          Then, a random number is compared with the ratio of sums of splitting functions taken at $z'$
          and their specific maximal value.
          \bea
          \#R \stackrel{?}{>} \frac{\sum\limits_{bc} P_{a\to bc}(z')}
                                        {\sum\limits_{bc} P_{a\to bc}^{\rm max}}\,,
          \eea
           where the $z'$ is accepted or rejected if the random number is larger or smaller than the ratio.
         }
\item{Having determined the decay kinematics by the $t'$, $z'$ the flavours of the outgoing 
          partons are selected according to the relative weight of the corresponding splitting functions at $z'$.
        }
\item{The outgoing partons are equipped with virtual masses themselves, starting from $t_a$.
          For each combination, Eqs.\,(\ref{Kincor}) and (\ref{Kincor11}) are applied to guarantee local four
          momentum conservation. If no combination of appropriate $t_b$ and $t_c$ can be found 
          respecting $t_i \ge t_0(i)$ and keeping $z'$ and the opening angle $\theta_{a\to bc}$ in the 
          allowed region, \APA returns to step 1 of this algorithm.  
          }
\item{The final task to be completed is to assign an azimuthal orientation to the
          decay plane with respect to the previous one. \APA provides two options, namely 
          \begin{itemize}
          \item{the uniform distribution of the relative angle $\phi$, or}
          \item{the inclusion of azimuthal correlations, \cite{Azimut},}
          \end{itemize}
          which can be chosen by the user.
         }
\end{enumerate}

\subsection{Fragmentation \label{physHad}}

After the parton shower has terminated at the cut-off virtuality $q_0^2$ the domain of long-distance interactions characterized by comparably low 
momentum transfer is reached. At this point QCD turns strong--interacting and non-perturbative effects take over their reign, converting the partons of 
perturbative QCD into the observable hadrons, a process which is called either fragmentation or hadronization. Since it is non--perturbative any 
traditional method of perturbative field theory meets with disaster and there is no approach derived from first principles to describe this 
process on a quantitative level. Consequently, the only way out is the construction of phenomenological models. 

Currently, \APA uses the fragmentation model provided by {\tt Pythia}, namely the Lund String--model \cite{And83}. Historically, the string 
hadronization scheme \cite{Art74,Bow81} was introduced as an alternative to the independent jet fragmentation scheme. The independent fragmentation 
scheme is the simplest and oldest model for translating partons into hadrons and was developped by Field and Feynman \cite{field}. Here, the hadronization 
of a $q\bar q$ pair is a recursive process starting with the generation of a secondary $q_1\bar q_1$ pair out of the vacuum. 
Then, the $q$ and $\bar q_1$ are combined into a meson. The procedure is iterated starting from the $\bar q q_1$ pair until the remaining energy of the 
corresponding left--overs falls below 
a cut-off. The production of the secondary quark pairs is modelled by the so--called fragmentation functions, yielding the probability distribution 
for a quark flavour $q$ to turn into a meson $M$ depending on the energy fraction $z=E_M/E_q$. Selecting the type of $M$ the flavour of the antiquark 
and thus the flavour and the remaining energy of the secondary quark pair is determined. In the independent fragmentation approach these functions 
are scale independent. The hadronization of a gluon can be incooperated by splitting the gluon into a $q\bar q$ pair. However, a shortcoming of the 
independent fragmentation scheme is, that the partons are treated  on--shell. This leads to a violation of four--momentum conservation, which has to 
be cured by rescaling the kinematics of the hadron ensemble, once the hadronization process has terminated.

In the string concept the $q\bar q$ pair is not independent any more but strongly correlated by a one-dimensional classical object,
the string. The string plays the role of the stretching colour field between the quarks and produces a potential between them which increases linearly
with their distance. The simplest $q\bar q$ configuration leads to the so-called yo-yo string and its classical evolution would result in an 
oscillation of the bound quark-antiquark pair. However, within a relativistic quantum mechanical system the energy can condense into the
production of a flavour neutral $q_1\bar{q}_1$ pair which screens the chromoelectric field. The resulting ensemble thus decouples into the two color 
neutral systems ($q\bar{q}_1$ and $q_1\bar{q}$), where each of them is subject to further dissociations into smaller systems.
So, hadronization is modelled as the break--up of a string in smaller ones, where each string hosts a $q_i\bar q_j$--pair at its endpoints,
which eventually are transformed into mesons or their resonances. Since the break--up of the strings into smaller ones is mediated by the
production of a secondary $q\bar q$ pair, the fragmentation functions encountered before come into play again, although in a slightly modified form.
In the Lund picture, the string break--up is interpreted in terms of tunneling phenomena, heavy masses are suppressed for the secondary quarks with 
ratios of roughly $u\bar u : d\bar d : s\bar s : c\bar c \approx 1 : 1 : 0.3 : 10^{-11}$. Additionally, the transverse momenta of the primary hadrons coming
into existence are chosen according to a Gaussian distribution with the width $\sigma_q$. This width is one major parameter of the hadronization, 
which has to be adjusted. The Lund fragmentation function reads

\bea
f(z) = z^{-1}(1-z)^a\exp{(-bm^2_{\perp}/z)}\,,
\eea

with $m_\perp$ the common transverse mass of the secondary $q\bar q$ pair determining the tunneling probability. Furthermore, the Lund fragmentation
function is left--right symmetric, i.~e. the results are independent on the choice of the starting point for the break--ups, quark or antiquark. Basically, the 
parameter $a$ could be flavour--dependent, while the parameter $b$ is not. However, phenomenologically there is no need to introduce different $a$'s. 
Thus the Lund fragmentation function has two parameters $a$ and $b$, which form the set of three major hadronization parameters to be set by the user in 
the file {\tt parameter.dat}. 

Again, in the simplest realization of the string model, the quark--antiquark pairs are transformed into mesons or their resonances with matching masses. 
More involved schemes like the Lund string allow for the incorporation
of baryons, too. For more details we refer the reader to the literature.
However, the hadrons themselves experience further decays of various types resulting in an ensemble with long--lived hadrons. 

In the string model gluons are incorporated as ``kinks'' on the string carrying finite energy and momentum. Rephrased in other words, unlike
the quarks the gluons are attached to two string pieces and thus their fragmentation is different from that of the quarks. Addionally, the kinks on the string 
also modify the dynamics. Hence, the one-dimensional yo-yo type description of the motion is not valid any more. Fortunately, covariant evolution equations
for kinky strings also exist. 

The string approach to hadronization has several advantages over the independent jet model. The basic assumptions of the 
string model seem to be in better agreement with the general ideas of QCD, on the lattice for example, ``flux tubes'' in quite a close analogy to the string have
been found. Furthermore, in the string model energy, momentum and flavour are conserved at each step of the fragmentation process, because
at each iteration (break--up) the whole system is considered. Last but not least, the results of Monte Carlo simulations  
are in far better agreement with experimental data.

%
%
%
%
%
%
%
\newpage
\section{Program Structure\label{PROG}}

In this section, we will discuss in some detail how the physics features outlined above manifest themselves
in the program \APA. We refer those of the users not interested in any internal details directly to 
Sec.\,\ref{inst}, where we list necessary prerequisites and steps to install and run \APA. 

However, since this program consists of roughly 8000 lines organized in 74 classes contained in
the {\tt C}--files plus slightly more than 2000 lines in the corresponding header files, and because there are
quite strong connections to the even larger program \AME, the description necessarily has some 
shortcuts. Nevertheless we hope, that the following subsections will provide any potential reader a sufficient
background for understanding the code. We start our presentation in Subsec.\,\ref{Strat} with a brief 
introduction into the basic strategies underlying \APA and the essential structures for their implementation. 
In Subsec.\,\ref{Steer} we describe, how \APA generates event samples and individual events.
The next part, Subsec.\,\ref{progXS}, is devoted to a discussion of the handling of the matrix elements, before we turn to 
the implementation of the parton shower in Subsec.\,\ref{progPS}. Finally, the issue of fragmentation within
\APA will be covered in Subsec.\,\ref{progHad}.

\subsection{Basic strategies and structures\label{Strat}}

In principle, \APA has its main focus on simulating the whole parton level of an event. Starting from the incoming beam particles, 
currently constrained to be an $e^+e^-$ pair, initial state radiation, hard scattering processes and the subsequent 
parton shower is covered. Then, after translating the parton ensemble appropriately into the {\tt HEPEVT}--block, some 
hadronization scheme is invoked, which at the moment is the Lund--string implemented in {\tt Pythia}. Consequently, the  
bulk of algorithms within \APA deals with the simulation of events on the parton level, the hadron level is covered via the
corresponding interface. Hence, our desciption of the basic strategies will focus on the parton level. 

\begin{figure}[h]
\centerline{\epsfxsize=14cm\epsffile{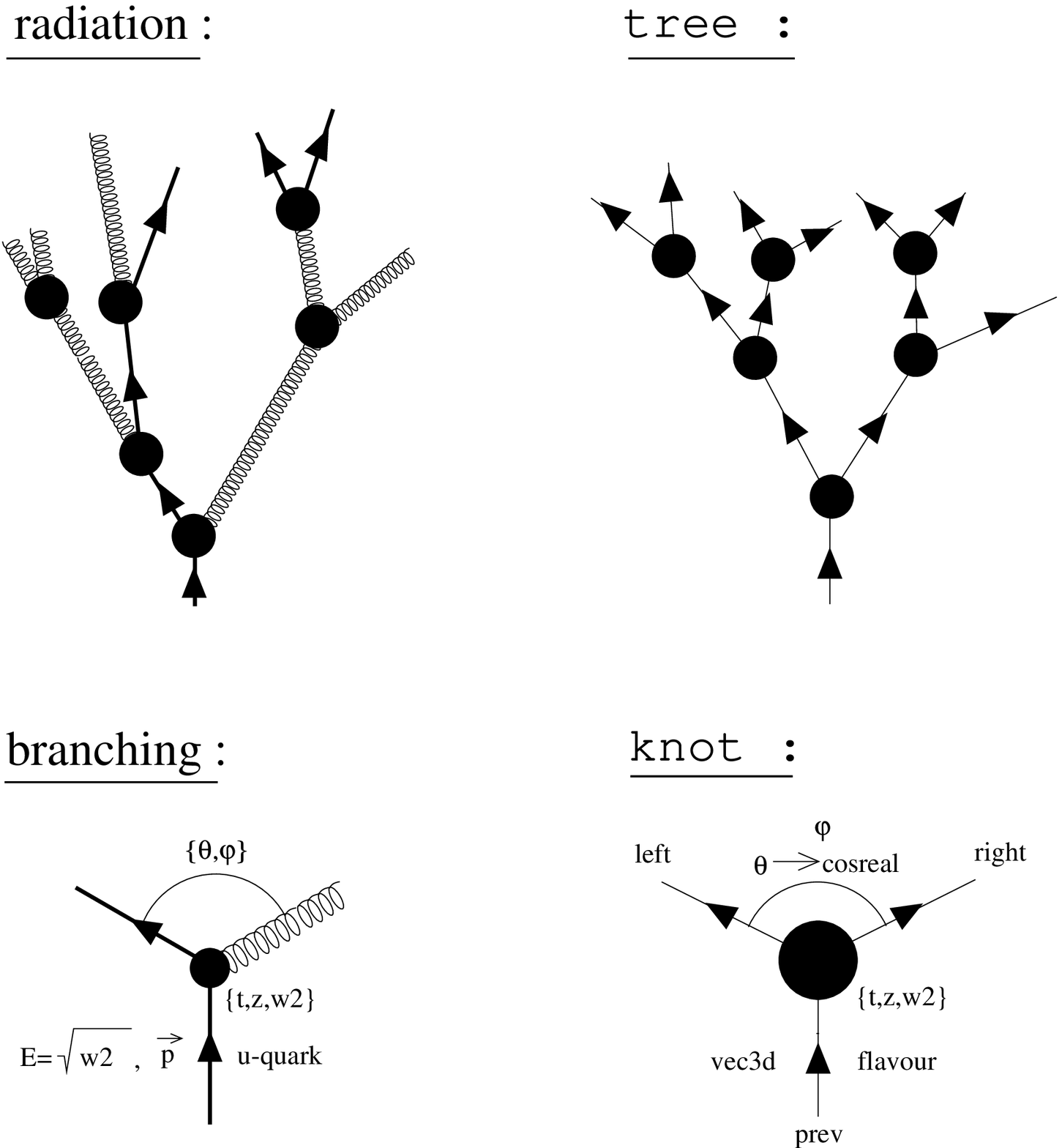}}
\caption{\label{mapps} Scetch of the mapping between radiation processes and the corresponding classes. 
                The full radiation pattern is identified as a chain of $1\to 2$ processes, a Markhov chain, which 
                translates into the class {\tt tree}. The basic building blocks, the binary decays, in turn are realized
                with {\tt knot}s. Thus a {\tt tree} contains a list of linked {\tt knot}s.} 
\end{figure}

The first observation underlying simulations in particle physics is, that the objects to be dealt with appear in two different
contexts. First, the particles can be classified according to their properties, i.~e. charges, masses and the like. In \APA
this information is contained primarily in the class {\tt flavour}, supplementing methods to define anti--particles or the link
between different numbering schemes for the particles. In contrast, the individual particles with their properties defined
in {\tt flavour} have to be tracked through a single event. The paradigma underlying \APA is to
\centerline{{\em define and treat partons within the event structure via their decays}.}
More specifically, the partons are dealt with by means of their $1\to 2$--decays. This is motivated by the following two 
observations:

\begin{enumerate}
\item{In the language of the leading logarithmic approximation, the radiation pattern of an event on the parton level
          reduces to a series of subsequent binary branchings, a Markhov--chain. Therefore, the basic building blocks
          of the parton shower can be identified easily with such $1\to 2$ decays, outgoing partons in this framework
          can be treated via ``non--existing'' $1\to 2$ decays.}
\item{Usually, the vertices encountered have either three or four external legs. However, within the Standard Model
          and its simpler extensions, the vertices with four legs can always be decomposed into the product of two 
          vertices with three legs and one propagator in between. In fact, this is the strategy employed within \AME.}
\end{enumerate}

Thus, the full radiation pattern of an event translates into a Markhov--chain of subsequent $1\to 2$ branchings, see Fig.\,\ref{mapps}.
This binary structure is recursive, since branchings follow each other. It is realized within the class
{\tt tree}, which technically contains a list of linked {\tt knot}s mirroring the basic building blocks, the branchings.
The {\tt knot}s harbour links to the {\tt prev}ious, the {\tt right} and the {\tt left} ones, allowing to climb up or down the 
tree by just following the pointers. In this framework partons entering fragmentation do obviously not experience any further decay 
and thus such ``dead ends'' are identified with {\tt knot}s with empty outgoing lines, i.~e. empty {\tt right}-- and {\tt left}--pointers.
To dwell a little longer on this issue, we would like to confront the branchings and the {\tt knots} with each other.
The branchings, for instance, are specified via :

\begin{enumerate}
\item{The three flavours, the incoming and the two outgoing ones, which in turn are incoming for the next splitting.
          The flavours define the splitting function of the decay, responsible for the $z$--spectrum of the decay.}
\item{The kinematical variables related to the decay, namely the virtual mass of the incoming particle, $t$ (or
          the $t=\theta E^2$--scale in MLLA), the energy fractions $z$ and $1-z$ of the two decay products and the
          azimuthal angle $\phi$. Together with the Energy $E$ and the three--momentum $\vec p$ of the decaying particle,
          the kinematics are fixed. For the inclusion of angular ordering ``by hand'', the opening angle $\theta$ then
          has to be compared with the previous one, $\theta_{\rm crit}$.}
\end{enumerate}
 
The {\tt knots} in full analogy include information about

\begin{enumerate}
\item{the predecessor and the two subsequent {\tt knot}s via pointers {\tt pref}, {\tt right}
          and {\tt left}, respectively, as well as the incoming {\tt flavour},}
\item{the kinematical parameters list above, namely {\tt t, ts, z, w2$=E^2$, cosreal$=\cos\theta$, crittheta$=\theta_{\rm crit}$}
          and $\phi$.}
\end{enumerate}

\begin{figure}[h]
\centerline{\epsfxsize=14cm\epsffile{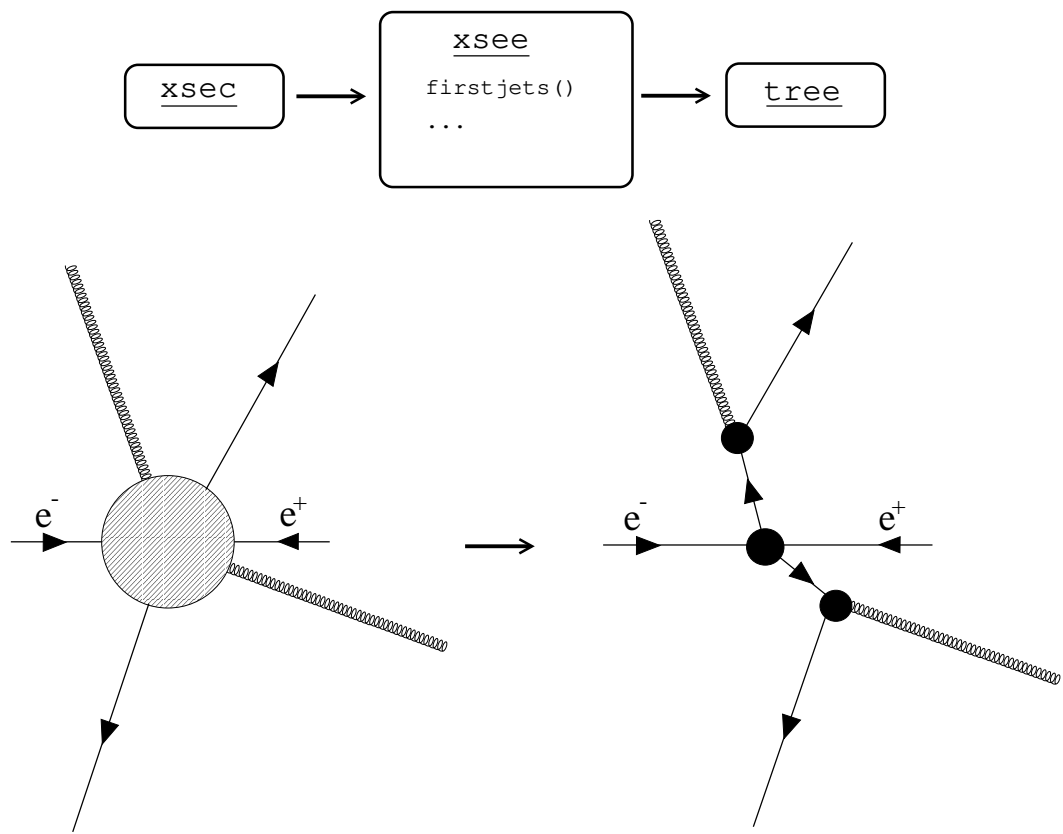}}
\caption{\label{mapxs} Scetch of the mapping between the hard cross sections as provided within {\tt xsec} and the
               further treatment of the final state via {\tt tree}. The translating class {\tt xsee} includes their derivatives, too.
               In fact these derivatives contain the interfaces to the matrix element generators yielding the cross sections.
               The interfaces are organized as a list within {\tt xsec}.} 
\end{figure}

Therefore, for their proper treatment within \APA, the final states stemming from the hard subprocess are translated into chains 
of subsequent $1\to 2$ decays, see Fig.\,\ref{mapxs}, before they experience their evolution down to the scales of fragmentation.
This is done with the help of the methods provided in the virtual class {\tt xsee} and their derivatives, providing interfaces
to the various matrix element generators. They in turn are organized
as a list within {\tt xsec}. In this context, we would like to 
stress, that the fragmentation scheme of {\tt Pythia} demands some specific information of the colour structure of an event. This 
is best formulated in some language relying on the parton shower approach incorporating the leading logarithmic approximation 
for multiple emission, too. Thus, already the final states produced by the matrix elements are translated into the {\tt tree}--structure,
independent of whether the subsequent parton shower models the jet--evolution or not.

Additional classes are frequently employed by other parts of the program. However, in most cases they are highly self--explanatory
and therefore do not demand any detailed discussion. They include {\tt
vec3d}, {\tt vec4d}, {\tt jetfinder}, etc..  The class {\tt random}
contains different random number generators, see \cite{RECIPES}.
Out of this group 
we would merely like to highlight some of the features of the class {\tt analyse} doing the event analysis. {\tt analyse} provides histogramms 
for some observables, namely

\begin{enumerate}
\item multiplicity, 
\item thrust, $C$-- and $D$--parameter, sphericity, aplanarity and rapidity with respect to the thrust--axis,
\item $p_\perp^{\rm in}$ and $p_\perp^{\rm out}$,
\item jet--broadening, 
\item $y_{5\to 4}$, $y_{4\to 3}$ and $y_{3\to 2}$, and
\item the four jet angles $\alpha_{34}$, $\chi_{BZ}$, $\phi_{KSW}$ and
$\theta_{NR}$ \cite{4JETth}.
\end{enumerate} 

Within the method {\tt init()}, all of these histogramms, which are classes themselves, are initialized. There, their individual number of bins and their range is 
defined, too. Hence, this is the place for eventual alterations. The methods {\tt fillanevent()} and {\tt drawallevents()} are responsible for filling in the data 
into the histogramms and for giving the final output. Some summarizing remarks can be found in the files {\tt allevent.dat}, other observables are to 
be found in corresponding {\tt .dat} files.

Note, however, that three classes {\tt analyse} analyze the events after the matrix elements, the parton shower and hadronization, respectively.
The corresponding output can be found in the subdirectories {\tt output/me.JOBNUMBER}, {\tt output/parton.JOBNUMBER}, and {\tt output/hadron.JOBNUMBER}.
The {\tt .dat} files are written out at least in steps of 10000 events.

\subsection{Generating events\label{Steer}}
\begin{figure}[h]
\centerline{\epsfxsize=14cm\epsffile{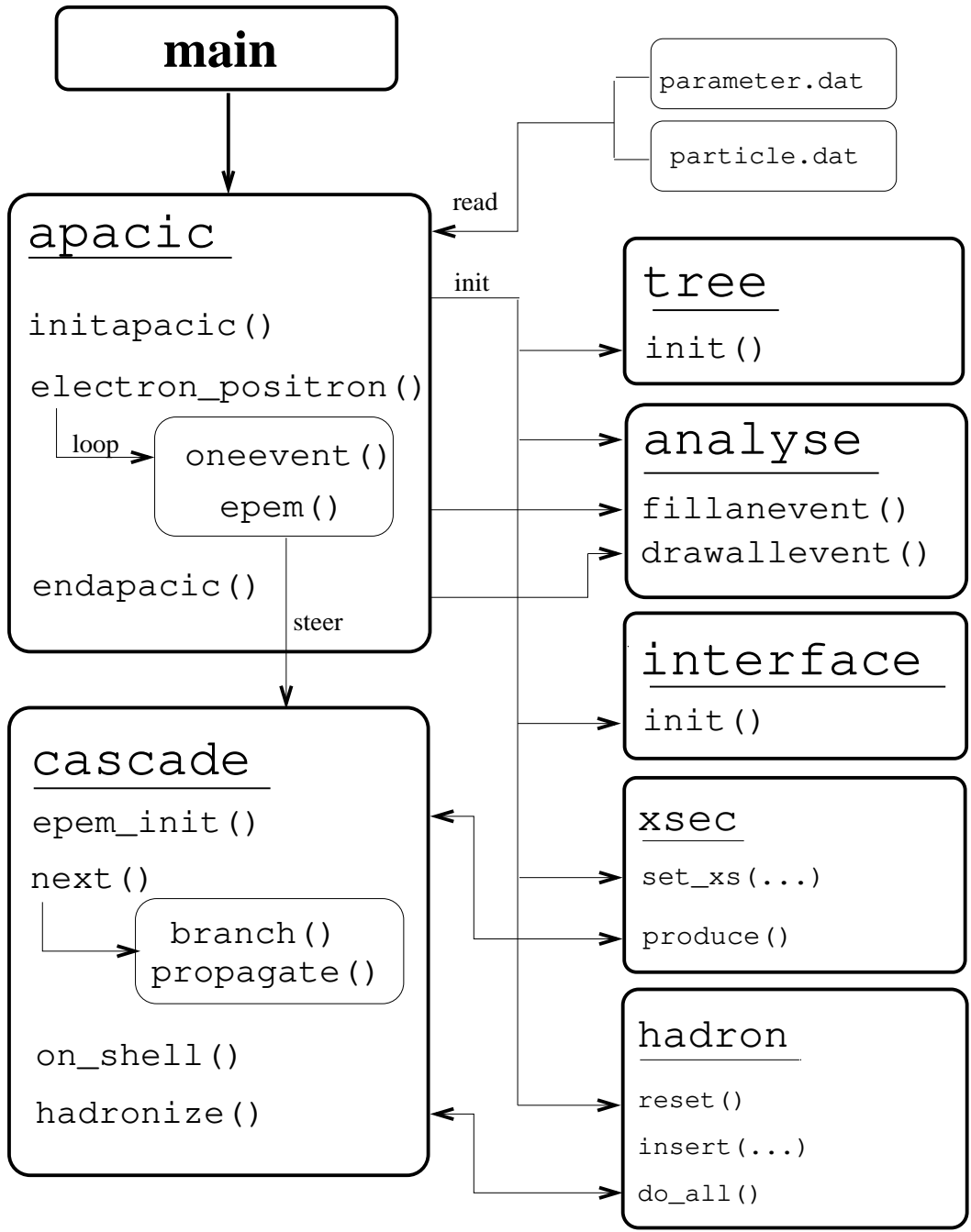}}
\caption{\label{steerfig1} Scetch of the interplay of the fundamental classes
    when running {\tt APACIC++}. Depicted are the two central classes of \APA, {\tt apacic}
    and {\tt cascade}, together with their most important methods and how they 
   cooperate. The communication and relationships of the two steering classes 
   with the other main parts of \APA are indicated, and the methods of the other 
   classes responsible for the contact with {\tt apacic} and {\tt cascade} are
   shown.}
\end{figure}
For the generation of events, \APA involves two central classes, namely {\tt apacic} and {\tt cascade}. In general terms, {\tt apacic} is 
the steering class responsible for the generation of event--samples and houses all the methods necessary to initialize and run a given 
number of events and to provide links to their analysis. Also, interfaces are included to link the other two event generators
available, namely {\tt jetset()} and {\tt herwig()}. However, we will not comment on them and focus on the running of \APA. 
When running {\tt apacic}, a loop over single events is performed within the class {\tt apacic}. In this loop individual events are initialized 
and simulated by means of the methods contained in {\tt cascade}. The hope behind this structure is, that it allows for a quick extension 
to other processes like e.~g. proton--proton collisions, and for a transparent link to other generators.

\subsubsection{\label{AllEv} Sample generation}
In \APA the central class responsible for the production of a sample of events and steering calls to each single event is {\tt apacic}. Its formal overhead 
is contained in {\tt main()}. Here, first {\tt apacic.init()} is called reading in the files {\tt particle.dat} and {\tt parameter.dat} and initializing 
particle information like charges and masses and the set of steering parameters and switches like coupling constants and the preferred shower 
scheme, respectively. Then, the c.~m.--energy squared, $s_{ee}$, and the number of events to be generated, $N_{ev}$ is transferred into the class
{\tt apacic}. Now the scene is set to chose the generator for the loop over events. \APA provides interfaces to {\tt Pythia} and {\tt Herwig}, 
but in the framework of this paper we want to concentrate on event generation by means of \APA.

Chosing this option, {\tt main()} calls {\tt apacic.electron\_positron()}. In this method, specific features for event generation via \APA are 
initialized with help of {\tt apacic.initapacic()}. Translated into the class structure of \APA, these include 
\begin{enumerate}
\item{the class {\tt tree} steering the parton shower with the method {\tt tree.init()}, see Subsecs.\,\ref{physPS} and \ref{progPS},}
\item{the class {\tt hadron} responsible for the subsequent hadronization by means of {\tt hadron.reset()}, see Subsecs.\,\ref{physHad} 
          and \ref{progHad},}
\item{the class {\tt interface} for the link to the corresponding
Fortran programs via {\tt interface.init()}, see SubSec.\,\ref{InterC}}
\item{the class {\tt analyse} for analyzing the events, and}
\item{the class {\tt xsec} handling the hard cross sections available with the more complicated call {\tt xsec.set\_xs(create\_xssum,$\dots$)}, 
          see Subsecs.\,\ref{physXS} and \ref{progXS}.}
\end{enumerate}
Now, within {\tt apacic.electron\_positron()} the loop over single events is performed, resulting in multiple calls of {\tt apacic.oneevent()}. The 
handling of the single events will be covered in the following Subsubsection, \ref{OneEv}. {\tt apacic.electron\_positron()} closes by calling the final 
analysis in {\tt apacic.endapacic()}. 
\subsubsection{\label{OneEv} Event generation with \APA}
The method {\tt apacic.oneevent()} envelopes the pre--event resetting of the classes {\tt tree} and {\tt hadron} responsible for the parton 
shower and the hadronization, respectively, and the steering method {\tt apacic.epem()}. This method provides the link to the class
{\tt cascade} and organizes the sequence of steps supplied there for the generation of single events. In chronological order, the methods 
of {\tt cascade} employed are:
\begin{enumerate}
\item{{\tt cascade.epem\_init()} determines the jet--configuration and eventually performs
          the jet--evolution by calling {\tt xsec.produce()}, see Subsec.\,\ref{progXS}. This method 
          returns a structure {\tt knot}, carrying all information about the subsequent chain of 
          branchings. Additionally, the momenta of the first two of the outgoing particles 
          are constructed. Their virtuality determines the distance they travel, before they decay.}
\item{{\tt cascade.next()} : A loop over all particles is performed, where each iteration is related to 
          some time measure. In each step {\tt cascade.branch()} determines, whether the particles 
          experience a decay or not. Note, that most of the characteristical parameters of the decays, 
          like kinematics and decay products, are already predefined. The corresponding information is
          contained in the {\tt tree} returned by {\tt xsec.produce()} via its root--{\tt knot} spanning it. After the branchings were
          performed, in each iteration the particles are propagated via {\tt cascade.propagate()}. The loop is 
          left, if no more branching takes place on the parton level.}
\item{{\tt cascade.onshell()} is finally applied to set all particles on their mass--shell under the 
          constraint of {\em global} four--momentum conservation. Thus, the parton ensemble is
          now prepared for hadronization. It should be noted, however, that since in the current version
          the Lund--string as provided by {\tt Jetset} takes care of hadronization, the corresponding
          {\tt HEPEVT}--block has to be filled in appropriately. This task is performed during each individual 
          branch described above via the method {\tt hadron.insert()}.}
\end{enumerate}

\subsection{Matrix elements\label{progXS}}

\begin{figure}[h]
\centerline{\epsfxsize=14cm\epsffile{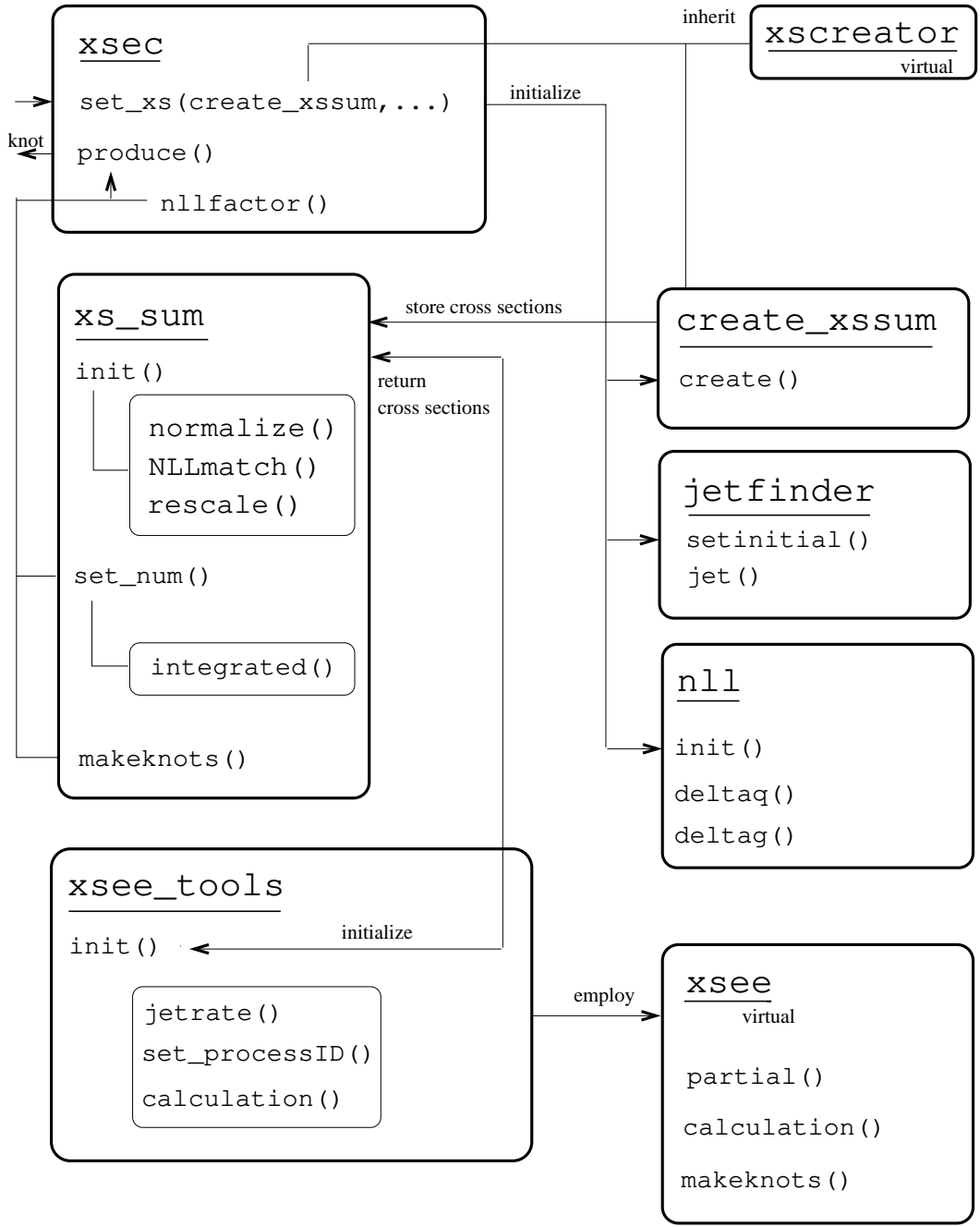}}
\caption{\label{steerfig2} Scetch of the interplay of classes governing the initialization,
   evaluation and running of cross sections for the hard subprocess.}
\end{figure}

The basic ideas behind the structure to be explained in the following are
\begin{itemize}
\item{to have only one class, {\tt xsec}, communicating with the steering classes {\tt apacic} and {\tt cascade},}
\item{to define standards for interfaces to a variety of matrix element generators in some virtual class {\tt xsee},}
\item{to organize the interfaces, i.~e. cross sections, in a list, {\tt xs\_sum} for easy access,}
\item{to keep any tools for the evaluation of total cross sections separate, {\tt xsee\_tools}.}
\end{itemize}
This leads naturally to a splitting in various classes, they and their mutual communication 
are depicted schematically in Fig.\,\ref{steerfig2}.

\subsubsection{Organization}

The class {\tt xsec} is the general steering class for the evaluation of matrix elements. 
It contains a list of interface classes in {\tt xs\_sum}, one for each channel under consideration. The interfaces 
represent the connection to the corresponding matrix element generators used and they
are derived from the virtual class {\tt xsee}. This class defines the minimal standard
of methods, each interface, i.~e. each individual matrix element, has to supplement when linked. 
The list of cross sections is organized by means of and contained in the class {\tt xs\_sum}.  
Only a few methods are employed within {\tt xsec}:
\begin{enumerate}
\item {\tt set\_xs} handles the initialization of the matrix elements and is called from {\tt apacic.initapacic()} with
           {\tt xscreator} and the incoming {\tt flavour} as arguments. Note, that the class {\tt xscreator} is a virtual class and 
           represents the mother for the two classes {\tt create\_xs} and {\tt create\_xssum},
           where only the latter is relevant in the following. The sequence of this initialization is :
           \begin{enumerate}
           \item The method {\tt xscreator.create()} initializes merely an array of interface classes, each 
                     derived from {\tt xsee}. This array is stored in
{\tt xs\_sum}. 
                     The incoming {\tt flavour} help to define the number of relevant channels to be initialized.
           \item The jetfinder algorithm needed for integration is initialized via the method {\tt jetfinder.setinitial()}.      
           \item Eventually the NLL--Sudakov form factors are calculated within {\tt nll.init()}. They are stored in  
                    form of a look-up table derived from the class {\tt fastfunc} and 
                    evaluated, i.~e. read via calling the methods {\tt nll.deltaq()} and {\tt nll.deltag()} for the quark
                    and gluon Sudakov form factor respectively. 
           \item The method {\tt xs\_sum.init()} initializes the interfaces to the matrix element generators 
using the array of step (a). It evaluates the total cross sections.
           \end{enumerate}
\item {\tt produce()} determines the specific final state of the individual event and a sample of momenta 
          distributed according to the differential matrix element plus some eventual extra weight. {\tt produce()} is called 
          by the method {\tt cascade.epem\_init()}. First,  a specific channel is chosen according to the jetrates by calling
          {\tt xs\_sum.set\_num()}. The determination of the corresponding momenta is accomplished
          with Monte Carlo methods according to the following procedure :
          \begin{enumerate}
          \item The maximum of the differential cross section under consideration is obtained from
	the method {\tt xs\_sum.maximum()}. 
          \item A sample of momenta as well as the appropriate differential cross section is determined via calling {\tt xs\_sum.partial()}. 
          \item The additional weight for the kinematical matching is evaluated according to the different options, where no weight at all,
	 the $\alpha_s$ or the Sudakov weight are at disposal. The latter is calculated with the method {\tt xsec.nllfactor()}.  
          \item The product of the extra weight and the differential cross section over the total maximum is compared with a random number. If it
	 is smaller the momentum sample is rejected and the procedure is repeated starting with step (b).
          \item The translation of the final state of the matrix element into the list of linked knots of the parton shower
	 and its further evolution are performed by calling the method {\tt xs\_sum.makeknots()}. If this step
                   fails, the procedure returns to point (b) as well.
          \end{enumerate} 
         The resulting list of linked knots representing the {\em full partonic stage of the event} is returned in the form of
         a link to the appropriate first {\tt knot}.
\item {\tt nllfactor} evaluates the weight for the Sudakov kinematical matching. The pre-calculated Sudakov form factors as well as
           the jetfinder for the determination of the jet resolution parameter $y_{\rm cut}$, {\tt jetfinder.jet()},  are prerequisites for
           this task and called accordingly.
\end{enumerate}

\subsubsection{Creating a {\tt xs\_sum}}

The class {\tt create\_xssum} produces a list of cross sections stored in the class {\tt xs\_sum}. This class appears 
when different channels, i.e. different final states for the same incoming particles, are included. However, 
in a typical {\tt APACIC++} run this is always the case. The only method of the virtual class {\tt xscreator} inheriting 
{\tt create\_xssum} is {\tt create}, where the incoming {\tt flavour}s represent the input and the accomplished 
list of cross sections the output. It is called by the method {\tt xsec.set\_xs()}. 
In {\tt create\_xssum.create()} first, the number of channels is specified. Then, according to
the number of jets and the possible outgoing flavours, channels are selected and the corresponding interface classes
are added. The specific choices depend on the parameter {\tt pa.jet()}, the switches connected with the selection of
matrix element generators (for instance {\tt sw.amegic()}) and the different models (for instance {\tt sw.QCD()} to
be used.  

\subsubsection{The list of cross sections, {\tt xs\_sum}}

The class {\tt xs\_sum} contains a list of interfaces to matrix element generators. It is responsible for all interactions with
them. In our approach an interface class is always derivated from the mother class {\tt xsee}, which defines the standard 
for implementing a new generator. Note, that for every process with fixed incoming and outgoing {\tt flavour}s the array 
includes a new interface class. The different methods of {\tt xs\_sum} fulfill the following tasks: 
\begin{enumerate}
\item {\tt init()} is used for the determination of the total cross sections and responsible for their proper normalization. It
           is called by {\tt xsec.set\_xs()}. First, the different interface classes will be initialized via {\tt xsee\_tools.init()}. The calculation of the 
           total section as well as the determination of the maximum of the differential cross section are the tasks of this method. Then, the cross
           sections have to be normalized via {\tt normalize()} to the appropriate inclusive $2\to 2$ process, for instance in the framework of 
           QCD to $\sigma_0(e^+\,e^-\to q\bar q)$. Now, the derived jetrates for different numbers of jets must be combined, which is not a unique 
           task. Three different schemes are at disposal, which are implemented in the method {\tt rescale()}.  
           One option then is to use resummed rates, which is achieved in the method {\tt NLLmatch()}.
\item {\tt normalize()} accounts for the proper normalization of the jetrates. The total cross sections for two outgoing particles are collected 
           depending on the outgoing quark flavour. Accordingly every cross section is normalized to the appropriate inclusive twojet rate.
\item {\tt rescale()} : The different schemes for combining jetrates with different numbers of jets are implemented in this method. Details
           can be found in the physics write-up,
Eqs.\,(\ref{Rdir}),(\ref{Rres1}) and (\ref{Rres2}).
\item {\tt NLLmatch()} calculates the resummed jetrates and matches them to the direct jetrates. First, for every number of jets the
           appropriate rate has to be determined. Then the resummed as well as the matched rate are determined with the method {\tt nll.calculate()}. 
           Finally, the different jetrates are rescaled according to the matched rate. Now, the twojet rate can be evaluated as one minus the sum of 
           the multijet rates.    
\item {\tt set\_num()} is called by the method {\tt xsec.produce()} and assigns the channel for the hard subprocess of the event 
          according to the jetrates. A marker ist set on the interface class of this channel, which is used for later evaluations with the specified 
          cross section.    
\end{enumerate}

During the event generation a number of additional methods are employed. They provide links to the interface class, 
which has been selected and marked in the routine {\tt set\_num()}. Typical methods contain the setting and reading
of the maximum of the differential cross section or the jetrate. In connection with the determination of a sample of momenta
the methods {\tt partial()} and {\tt get\_ycut()} are employed. They return the differential cross section and the minimal jet 
resolution parameter $y_{\rm cut}$ of the jet constellation, respectively. The method {\tt makeknots()} is responsible 
for the combination of the chosen matrix element with the parton shower evolution.  

\subsubsection{Integration of ME's}

The integration of the matrix elements resulting in the total cross sections is governed by the class {\tt xsee\_tools}. The following
methods are used for calculating, storing and reading in results.
\begin{enumerate}
\item {\tt init()} is the only method called from outside this class, i.e. from the method {\tt xs\_sum.init()}. 
          Its first step consists in the determination of the appropriate power of $\alpha_s$. This is, because inside the matrix 
          element generators $\alpha_s(s)$ is used, and correspondingly a factor of $(\alpha_s(\kappa_S s)/\alpha_s(s))^{N_{\rm jet}-2}$ 
          has to be multiplied to every total cross section for consistency reasons when including scalefactors $\kappa_S$.
          Note, that this procedure holds in the framework of pure QCD only. In the second step the jetfinder and the interface to
          the matrix element are initialized. Now the total cross section can be calculated by means of the method {\tt jetrate()}, 
          which yields the maximum of the differential cross section, too. In the last step the resulting values for the
          maximum and the total cross section are set in the interface class.    
\item {\tt jetrate()} maintains the calculation of the total cross section, which contains not only the evaluation but
          also the storage of the results for later use. For this purpose, first every process is equiped with an ID in {\tt set\_processID()},
          which depends on the specific final state. Then the directory {\tt me} is searched for the corresponding file. In case it is found,
          a simple read--in by means of the method {\tt input()} finishes this routine. Otherwise the method {\tt calculation()} 
          determines the total cross section as well as the maximum of the differential cross section. Finally, both are stored in a
          table with the corresponding values in dependence on the jet resolution parameter $y_{\rm cut}$. 
\item {\tt calculation()} handles the determination of the total cross sections employing the following steps :
          \begin{enumerate}
          \item The look--up table for the total cross sections is initialized with the method {\tt histofunc.reset()}. 
          \item The phase space generator is initialized via its constructor {\tt psgen()}. Note, that this generator is lend from the matrix
                    element generator \AME. A description of the
different modes, which are the integration of the matrix element 
                    with {\tt Rambo}\cite{RAMBO} and with multichannel
methods\cite{Weightopt}, can be found in \cite{AMEGIC}. The corresponding method used 
                    for the generation of the phase space is given by {\tt sw.multichannel()}.
           \item At this stage a loop over the corresponding Monte Carlo points is performed. In every step a point in
                     phase space together with its weight will be generated with the method {\tt psgen.partial()}. Then the minimal 
                     $y$ of the four vectors is calculated with {\tt jetfinder.y\_jettest()}. The value of the differential cross section, obtained 
                     from {\tt xsee.partial()} and multiplied with the proper weight of this phase space point is stored in the look-up table 
                     via {\tt histofunc.insert()}. In case the multichannel-method was chosen the loop ends with an optimization
                     step in {\tt psgen.optimate()}.  
           \item Finally, the look-up table is stored by means of the method {\tt histofunc.output()}. 
           \end{enumerate}
\end{enumerate}  

\subsubsection{Interfaces\label{XSInt}}

\begin{figure}[h]
\centerline{\epsfxsize=14cm\epsffile{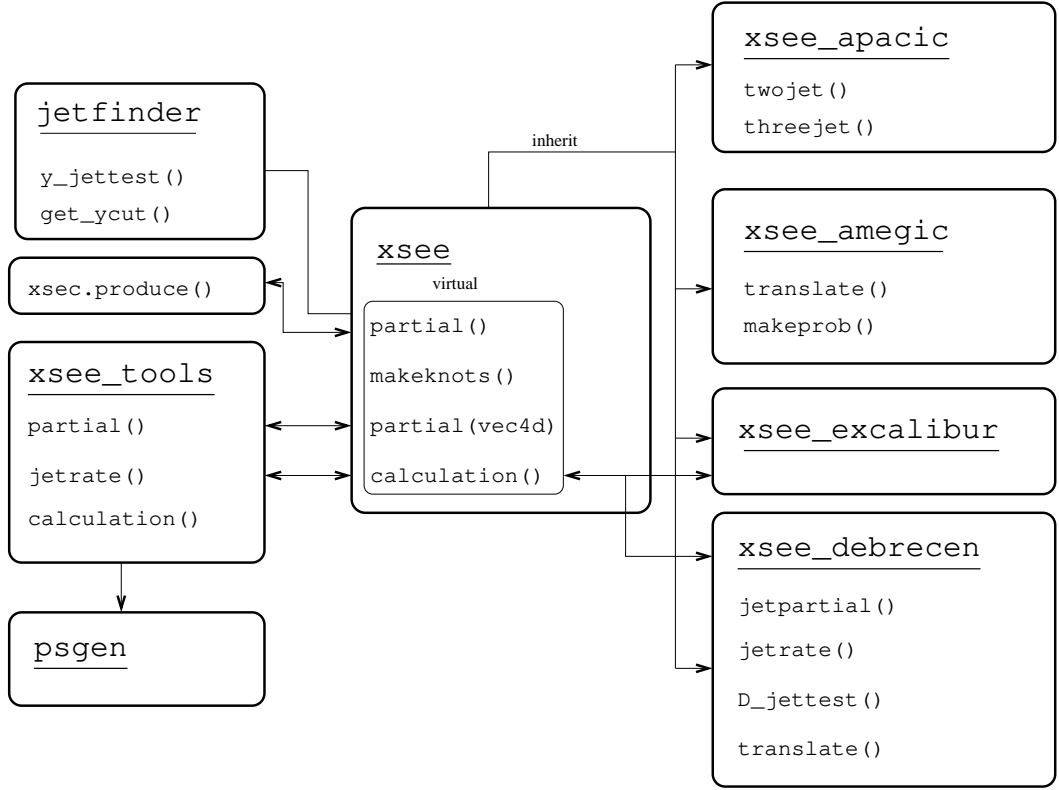}}
\caption{\label{interfig} Scetch of the various interface classes available within \APA for the connection to
the matrix element generators, which can be linked. Included are some of the communication lines
with other classes handling cross sections, namely {\tt xsec} and {\tt xsee\_tools}.}
\end{figure}

All interface classes are derivated from the class {\tt xsee}, see Fig.\,\ref{interfig}. It defines the standards for communicating with any 
matrix element generator used. The class is purely virtual, i.e. it has no genuine method. However, since most of the methods in all interfaces have 
the same purpose, they are described at this stage. 

\begin{enumerate}
\item {\tt partial()} determines a sample of momenta, the appropriate weight in phase space and the 
           differential cross section. Performing these tasks the methods of the considered matrix element generator are used. 
           The mininal jet resolution parameter is determined by the method {\tt jetfinder.y\_jettest()} and accessed via {\tt get\_ycut()}.
           {\tt partial()} is used by the method {\tt xsec.produce()} and returns the differential cross section multiplied with the 
           phase space weight.
\item {\tt partial(vec4d)} in contrast returns the differential cross section for a given sample of momenta, which is represented 
           as a list of 4-vectors ({\tt vec4d}). It is called by the method {\tt xsee\_tools.partial()}.   
\item {\tt calculation()} : A calculation of the total cross section inside the matrix element generator is implemented only in two
           interfaces. The reasons are in the first case, that a Next-to-Leading order calculation is performed ({\tt DEBRECEN}) and in 
           the second case, that a multichannel approach during the integration is used ({\tt EXCALIBUR}). Obviously, in both cases
           specific features of the corresponding calculation or method had to be encoded. In the latter case an additional issue to be 
           dealt with was transformation of the momenta into the internal structure. When necessary, this routine 
           is called from {\tt xsee\_tools.jetrate()} instead of the generell integration routine {\tt xsee\_tools.calculation()}. 
\item {\tt makeknots()} : The combination of the matrix elements and the parton shower strongly depends on the generator under 
          consideration and its internal structure. Consequently, a non--general method {\tt makeknots()} is needed for this task, 
          where the different approaches of reconstructing a parton shower history are implemented, see Subsec.\,\ref{physPS}.
\end{enumerate}

Additional methods for exchanging informations between the interfaces and the class {\tt xs\_sum} are provided. Most of them 
are self--explanatory, therefore we will not discuss them in detail. Consider as examples the methods {\tt njet()}, {\tt maximum()},  
and {\tt integrated()} yielding the number of jets, the maximum of the differential cross section and the total cross section,
respectively. 

\begin{table*}[h]
\begin{center}
\begin{tabular}{|l|lcccll|}
\hline
Interface Class & Number & QCD & EW & Mass & Matching & Other Features\\
Name & of Jets &&&& Options &\\
\hline
xsee\_apacic & 2,3 & X & - & X &0& none\\
xsee\_amegic & 2,3,4,5 & X & X & X &0,1,2,3& + Higgs\\
xsee\_debrecen & 3,4,5 & X & - & - &0& + NLO for 3,4 jets\\
xsee\_excalibur & 4 & X & X & - &2,3& 4 fermions only\\
\hline
\end{tabular}
\caption{\label{gentab1} Matrix Element Generators interfaced to {\tt APACIC++}}
\end{center}
\end{table*}

The interfaces to the four different matrix element generators, namely the internal generator of {\tt APACIC++}, \AME, 
{\tt DEBRECEN} and {\tt EXCALIBUR}, as well as some specific features are listed in table \ref{gentab1}. As already explained in 
Subsec.\,\ref{physXSPS}, \APA provides different options for the determination of the relative probabilities connected to the 
colour structure of a given final state produced by some matrix element. Therefore, within the combination procedure of
matrix elements and the parton shower, these different approaches for reconstructing a parton shower history are
reflected in different algorithms with corresponding methods, namely 
\begin{enumerate}
\item[0:] The parton shower history will be reconstructed by means of the method {\tt tree.histjets()}. Employing this method, the
               matrix element generator is not used for the determination of the relative probabilities of the different colour
               configurations. Instead, the probabilities are calculated in the parton shower oriented picture.   
\item[1:] This option exist for the interface {\tt xsee\_amegic} only and is performed within the method {\tt makeknots()}. 
                \AME constructs the Feynman amplitudes via binary trees of linked points, which is in striking analogy to
               the {\tt tree}--structure of \APA. Thus, this method merely ``translates'' the matrix element--points, which include all
               kinematical information needed, into the linked {\tt knots} of {\tt tree}. Consequently, the calculation of the parton shower 
               probability like the one encountered in Eq.\,(\ref{partprop}) is straightforward.
\item[2:] The probability of a parton history is proportional to the appropriate matrix element squared, see the first of 
               Eqs.\,(\ref{dirprop}). Hence, it has to be calculated with the matrix element generator used. 
\item[3:] When including interference effects between the different matrix elements in the spirit of the second of Eqs.\,(\ref{dirprop}), 
               the probability has to be evaluated inside the matrix element generator, too. 
\end{enumerate}

Every interface class contains some specific methods beyond the standard methods defined within the virtual mother 
class {\tt xsee}. These additional methods are neccessary for the appropriate connection to the matrix element
generator. Furthermore, every class provides a different implementation of the method {\tt makeknots()} in accordance with
the different options used for the determination of the relative probabilities highlighted above. In the following, we briefly
outline the additional methods, and we comment on the details related with {\tt makeknots} for each of the interface classes.
\begin{enumerate}
\item {\tt xsee\_apacic} : Since all methods for the calculation of the cross sections are contained 
                                          within the interface class itself, the tasks and methods are :
          \begin{enumerate}
          \item The calculation of the differential cross sections in {\tt twojet()} and {\tt threejet()} for the processes 
                    $e^+\,e^-\to q\bar q$ and $e^+\,e^-\to qg\bar q$, respectively. 
          \item The determination of the total cross sections in {\tt r\_qq\_bar()}, {\tt sigmaWW()} and {\tt sigmaZZ()} for the 
                   electron positron annihilation into quark, $W$--boson, and $Z$--boson pairs.
          \end{enumerate}
          The reconstruction of the colour configuration within {\tt makeknots} relies on {\tt tree.histjets()}.
\item {\tt xsee\_amegic} : Interfacing the matrix element generator \AME all four different options for the reconstruction of
                                           the colour configurations and their relative probabilities are available. The first one corresponds to the 
                                           parton shower oriented approach and employs the method {\tt tree.histjets()} for the histories and their 
                                           probabilities. For the three other options the probabilities of the colour configurations are determined as 
                                           follows: 
         \begin{enumerate}
         \item[1:] The translation of the internal representation of the Feynman diagrams within \AME into probabilities is 
                        performed via the method {\tt makeprob()}. In recursive steps the tree of linked points is used to evaluate the 
                        parton shower oriented probability. In each point representing a $1\to 2$--vertex, the energy fraction $z$ and the virtual
                        mass of the incoming as well as the flavours of the outgoing partons are used. The appropriate splitting function 
                        is taken into account by {\tt timebranch.set\_ds()} utilizing the two outgoing flavours.  The value of the splitting function 
                        is calculated with {\tt (timebranch.dsp).differ()}. After combining it with the virtual mass of the incoming particle, 
                        the contribution of this {\tt point} to the probability is determined. The method {\tt makeprob()}  is called recursively
                        with the {\tt left}-- and {\tt right}--pointers of the structure {\tt point} until the calculation terminates with the
                        outgoing partons.
         \item[2,3:] The probabilities are calculated during the evaluation of the matrix elements within \AME.  
         \end{enumerate}  
         Having at hand the probabilities for each colour configuration, one of them is choosen accordingly. In the next step 
         eventually the linked {\tt points} of \AME are translated into the linked knots of \APA. The procedure employed is the 
         same for the last three options above. Again, the corresponding method {\tt translate()} employs a recursive structure, 
         where in every step a {\tt point} is translated into a {\tt knot}, the pointers to the left and the right point are
         translated into links for the related new {\tt knots} initialized with {\tt tree.newk()}.
\item {\tt xsee\_debrecen} : Since the calculation of Next-to-Leading order cross sections involves algorithms, which are a priori not included 
          in the integration routines contained in {\tt xsee\_tools}, specifc methods are neccessary for this task:  
          \begin{enumerate}  
          \item {\tt jetrate()} : Three different parts for the calculation of NLO cross sections must be considered. Therefore, this 
                                            method can be called from the routine {\tt calculation()} which leaves the evaluation of cross sections
                                            to the matrix element generator. Accordingly, three different terms, related to the born term, the loop 
                                            corrections and the real corrections due to additional legs, have to be added. The distinction between 
                                            these parts is shifted to subsidiary methods. However, the calculation with the typical loop over the 
                                            events is performed in the usual manner. For the evaluation of the differential cross sections the 
                                            method {\tt jetpartial()} is employed.    
          \item {\tt jetpartial()} : Typically, the random construction of  momenta is the first step in the calculation of a differential cross
                                                section via Monte Carlo methods. Here, difficulties arise from the different dimensions of the phase 
                                                space for the different parts. Therefore the phase space for $N+1$ particles is created out of the $N$ 
                                                particle phase space. After the calculation of the minimal jet resolution parameter with {\tt D\_jettest()} 
                                                the appropriate differential cross section is determined.      
          \item {\tt D\_jettest()} evaluates the minimal jet resolution parameter for a given sample of momenta and a 
                                                 pre-defined number of jets. It mirrors the methods of the class {\tt jetfinder}, but with the slightly 
                                                 different description of 4-vectors used in {\tt Debrecen}. 
          \item {\tt translate()} transforms a 4-vector from the {\tt APACIC++} ({\tt vec4d}) to the {\tt Debrecen} 
                                                ({\tt LorentzVector}) convention.
          \end{enumerate}
          The calculation of the probabilities for the colour configurations within the combination procedure is performed with 
          {\tt tree.histjets()} from {\tt makeknots()}.   
\item {\tt xsee\_excalibur} :  Like in \AME the different probabilities are calculated within the program during the 
                                               determination of the differential cross section. The only task left then is the translation of the structure of 
                                               Feynman diagrams into the list of linked knots, which is achieved in the method {\tt makeknots()} of the
                                                class {\tt xsee\_excalibur}. This method employs the fact, that {\tt Excalibur} specifies external legs of
                                                predefined topologies. Employing the method {\tt tree.construc()} the knots can then be translated into 
                                                a binary tree.
\end{enumerate}

\subsection{Parton shower\label{progPS}}

\begin{figure}[h]
\centerline{\epsfxsize=14cm\epsffile{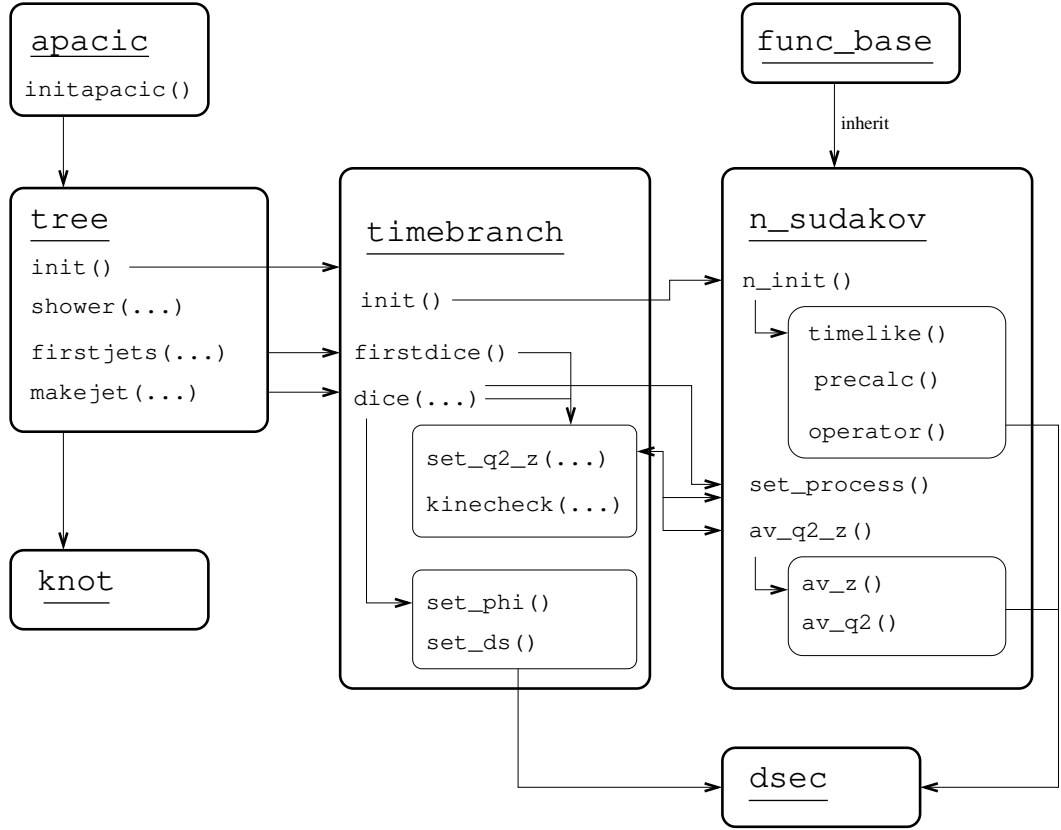}}
\caption{\label{psorgfig} Scetch of the interplay of classes governing the parton shower. The central class
{\tt tree} organizes the parton shower as a binary tree of subsequent parton decays each represented by a knot.
{\tt timebranch} in turn governs the individual parton branchings with help of the Sudakov form factors and
splitting functions. The corresponding methods for the handling of Sudakov form factors and splitting
Functions can be found in {\tt n\_sudakov} and {\tt dsec}, respectively.}
\end{figure}

\subsubsection{Organization - {\tt tree}}

The class {\tt tree} is the central steering class for the evolution of partons from higher to lower scales via multiple binary 
decays. It houses the organization of the parton shower and parts of the combination of matrix elements with the parton shower. 

Basically, {\tt tree} contains a list of knots, which properly linked represent a binary tree, i.~e. a chain of branchings
$a\to bc$, where $b$ and $c$ themselves eventually branch. This mirrors the physical structure of the parton shower, which is
constructed by means of subsequent parton splittings. A number of methods are dedicated merely to the handling of this 
list of (linked) knots:          

\begin{enumerate}
\item {\tt newk()} returns the link to a new knot. During extensions of the tree, i.~e. until all partons have reached
           the infrared cut--off $q_0^2$ of the parton shower this method is frequently used.
\item {\tt add()} : With this method a new tree can be added to the actual one. Note, that as a result the two list of knots are only 
           formally connected. Any additional, more specific link between them has to be established by hand.   
\item {\tt operator+()} : Two trees can be added to form a new tree. The same remarks as for the method {\tt add} hold.
\end{enumerate}

The parton shower part of the class {\tt tree} is organized as follows: 
\begin{enumerate}
\item {\tt init()} is the main routine for initializing the parton shower. It is called from {\tt apacic.initapacic()}. First, the class
          timebranch is activated via its constructor {\tt timebranch()} and initialized with {\tt timebranch.init()}. Now, the Sudakov 
          form factors can be calculated. For this, the method {\tt timebranch.timelike()} is responsible. Either it evaluates or
          it reads in the form factors, which, when already calculated, are stored in a look-up table accessible via the class {\tt fastfunc}.     
\item {\tt shower()} is the main routine for the organization of the parton shower. The input is a list of knots with already 
          established links. Therefore {\tt shower()} is always called after the selection of a parton shower history for the appropriate 
          matrix element, i.~e. after succesful translation of the matrix elements final state into linked knots. Note, that in this case the
          partons are still on-shell. Hence, at first the method {\tt firstjets()} is called supplying the partons with virtual masses. 
\item Only then {\tt makejet()} is called, which implements the parton shower evolution in a recursive manner. The arguments are a 
         mother, a grandmother (``granny'') and two daughter {\tt knot}s. Note, that without azimuthal correlations included, the parton shower needed 
         a mother knot only. However, {\tt makejet()} returns a knot with established links. 

        Now the sequence of one individual branching within {\tt tree()} is
        \begin{enumerate}
        \item The value of the link to the mother knot is checked. If it is zero, the parton shower stops.  
        \item A link between the mother and the granny knot is established through the {\tt prev} pointer of the mother.
        \item The two new daughters are initialized and connected to a knot with the method {\tt newk()}.  
        \item The two daughter knots are filled by means of the method {\tt timebranch.dice()} and the kinematics of the mother knot is 
                 changed accordingly. Eventually, all four knots, i.e. granny, mother and the two daughters, are employed for this task.
        \item The pointers to the left and right knots of the mother are set by calling the method {\tt makejet()} recursively. In this step
                  the appropriate daughters become mother and then, of course, the mother knot itself transforms into the granny.   
         \item Finally, the mother knot with all links will be returned.
         \end{enumerate}
        Note, that this procedure terminates in each branch of the tree in case of a zero link. The corresponding decision is made 
        within the method {\tt timebranch.dice()}, which returns a zero for both daughters, if no further branching is possible.   
\end{enumerate}

Some parts employed for combining the parton shower and the matrix elements are contained within {\tt tree}. 
These are the reconstruction of the colour configuration in the parton shower approach as well as the determination
of virtual masses for the outgoing particles of the matrix element.  The methods employed for these two purposes are:

\begin{enumerate}
\item {\tt histjets()} needs as input a list of momenta and the corresponding flavours. It yields as result the fully
            reconstructed parton shower history. The various colour configurations including their relative probabilities
            and the selection of one of them is contained in {\tt topology()}. Having chosen the history, however,
            the corresponding binary tree experiences the subsequent parton shower by the method {\tt shower()}.
\item {\tt nohistjets()} is used, if the colour configuration was already selected and is given by a list of knots. 
          With {\tt construc()} the knots are then connected and filled into the {\tt tree}. Again, {\tt shower()} performs 
          the further parton shower evolution.
\item {\tt topology()} reconstructs the parton shower histories related to each colour configuration and selects 
          one of them.  The following procedure is applied :
          \begin{enumerate}
          \item The maximal number of parton histories and the number of knots in each history are determined. 
                    Then the last $N_{\rm jet}$ knots representing outgoing particles are filled in all permutations with the outgoing 
                    particles, i.~e. with the flavours and momenta obtained via {\tt histjets()}.
          \item Starting from this representation of the final state, the method {\tt recstep()} recursively constructs the knotlists of 
                    the different histories. In each iteration, the contribution to the probabilities is calculated as well.
          \item One filled history is chosen according to the determined probabilities. 
          \item The selected history, which consists of a knotlist, is filled into the structure of {\tt tree} with {\tt construc()}. The 
                    links between the different knots are established in this step as well.  
          \item A link to the root of the new {\tt tree} is returned.
\end{enumerate}    
\item {\tt recstep()} determines recursively all possible histories through successive recombination of two partons. Consequently, 
          the problem reduces in every step from the reconstruction of a $n$ parton history to a $n-1$ parton history. The algorithm 
          ends with the final recombination of two partons to the initial $\gamma^*/Z$ stemming from the $e^+\,e^-$ annihilation. 
          In every step the contribution of the corresponding branching to the overall probability is calculated. For this purpose
          the splitting functions of {\tt dsec} need to be employed. The reconstruction of all parton histories is performed as follows:
          \begin{enumerate}
          \item The termination of the recursion is achieved when only two partons remain. They are combined into the root knot.    
          \item If more then two partons appear, a loop over their number is started.  
          \item With the method {\tt timebranch.set\_ds()} a check, if a parton and his neighbour can be combined, is enforced. 
                   In the progress of this check the splitting function is initialized, too. Passing this test the two partons are combined into 
                   one, otherwise the procedure continues at point (g).  
          \item The probability for this branching is calculated by multiplying the value of the splitting function from 
                   {\tt (timebranch.dsp).differ()} with the propagator of the new mother, see Eq.\,(\ref{partprop}). 
          \item The new knot is equiped with the informations gained from the two partons. Then it is filled into the histories.
                    Each of them is represented by a knotlist. Note, that for every succesful combination of partons a number of possible 
                    histories arise, which have to be traced.
          \item A call to {\tt recstep()} starts the next recursion step.
          \item The two partons can not be connected (for instance, if both of them are quarks). Consequently, all parton
                   histories which rely on this combination have to be deleted. This is achieved by setting the probabilities to zero.
          \end{enumerate}
\item {\tt construc()} : This recursive method fills a list of unlinked knots into the structure of {\tt tree}, i.e. it provides the
         links between the different knots. Therefore it starts with the root knot as the first mother and searches for the two
         daughter knots. The energy of the mother together with the energy fractions $z$ for the daughters are used to look for 
         the matching daughter energies. The knots are linked and the procedure starts again with the two daughters as mothers.
\item {\tt firstjets()} is exclusively called by {\tt shower()} and provides the outgoing partons of the matrix element with their
         virtual masses by employing {\tt timebranch.firstdice()}. The second task of {\tt firstjets()} is the determination of the 
         azimuthal angles of the branching planes within the matrix element. This information is mandatory for the proper
         evaluation of the azimuthal angles in subsequent branchings.   
\end{enumerate}

\subsubsection{Filling a knot - {\tt timebranch}}

The class {\tt timebranch} handles individual splittings, i.e. it is responsible for the decision whether a parton decay happens or not
and for filling the corresponding kinematical variables into the appropriate knot. Technically it is a derivative of the class 
{\tt n\_sudakov}. The important methods are {\tt firstdice()} and {\tt dice()} for the first and the subsequent parton decays, respectively. 
{\tt timebranch} is exclusively called by methods of the class {\tt tree} and utilizes the classes {\tt n\_sudakov} and {\tt dsec} 
for the calculation of the Sudakov form factors and the splitting functions ($P(z)$) respectively. Its methods are

\begin{enumerate}
\item {\tt init()} is called by the method {\tt tree.init()}. It initializes the internal jet clustering scheme with {\tt jetfinder.setinitial()} and the
          Sudakov form factors with {\tt n\_sudakov.n\_init()}. For the determination of the coefficients for the azimuthal correlations 
          between different branching planes the splitting functions are needed. They are related to the different decays occuring inside the parton shower, 
          namely $g\to gg$, $g\to q\bar q$,  $q\to gq$,  and $q\to qg$. The splitting functions are created via the class 
          {\tt create\_ds} and stored into the appropriate {\tt dsec} with the method {\tt dsec.set\_ds()}. Note, that this construction 
          parallels the handling of the cross sections in {\tt xsec} and the adjacent classes. 
\item {\tt dice()} is the main routine for one branching. A granny, a mother and two daughter knots are the incoming arguments, 
          with only the granny knot staying unaltered at all. The determination of the branch, i.~e. its possibility and 
          kinematics, is performed in the following steps:
          \begin{enumerate}
          \item The process is initiated with the method {\tt n\_sudakov.set\_process()}. Arguments are the ordering scheme, i.e. virtuality
                   or angular ordering, and the flavour of the mother.
          \item The flavours of the daugthers are determined with the method {\tt n\_sudakov.out()} utilizing the energy fraction $z$ and 
                    the virtuality $t$ of the mother knot, which have already been selected in the previous decay.
           \item The energies of the daughters are set and the starting values for the evaluation of their virtual masses are
                    identified with the mothers virtuality. Now, the main loop starts with the daughter virtualities decreased in each step
                    until a kinematical allowed constellation is achieved. 
           \item In each iteration of the loop, the virtual mass of only one daughter is decreased. The corresponding daughter is selected 
                    according to the bigger ratio of virtuality and energy. The method {\tt set\_q2\_z()} with the appropriate daughter 
                    knot as argument, performs this step resulting in a new pair of virtual mass and energy fraction, $\{t,z\}$, for the daughters
                    subsequent decay. 
           \item A first simple kinematical test checks whether the sum of the square roots of the daughter virtualities are smaller
                    than the square root of the mothers virtual mass. In case of a failure, the procedure continues with (i).
           \item The energy fraction $z$ of the mother is corrected according to Eq. \,(\ref{Kincor11}).
           \item The main check for the consistency of the kinematical variables, obtained in the last step, is performed with the 
                     method {\tt kinecheck()}. Again, a failure leads to point (i). 
           \item If the combination of the two pairs $\{t,z\}$ of the daughters is accepted, the variables can be filled into the appropriate
                    knots. The determination of the azimuthal angle between the plane of the two daughters in respect to the plane spanned by
                    the vectors of  granny and mother is performed with {\tt set\_phi()}. This step finalizes {\tt dice()}.    
           \item If any of the kinematical checks fails, the sequence continues at this point. In case the daughters have already reached 
                    the cut-off virtuality $q_0^2$ with still no kinematical fit, an additional decay of the mother does not happen.
                    Then, the mothers virtual mass by definition equals $\mbox{\rm min}\{m_m^2,q_0^2\}$ and the daughter knots are
                    set to zero. If the daughter knots have not yet reached the cut--off $q_0^2$, the procedure continues with step (d).
            \end{enumerate} 
\item {\tt firstdice()} determines the first virtualities, with partons stemming directly from the matrix element. The difference 
                     to {\tt dice()} manifests itself physically in the possible branching of a parton into a fixed propagator and an 
                     outgoing parton, see Subsec.\,\ref{physXSPS}. Therefore, in such cases the kinematics of one daughter is fixed 
                     a priori resulting in the usage of Eq.\,(\ref{Kincor10}) instead of Eq.\,(\ref{Kincor11}). However, checks of the kinematics 
                     with {\tt kinecheck()} and the sequence of steps remain unaltered.
\item {\tt kinecheck()} : Regarding one branch the kinematical checks to be applied are :
          \begin{enumerate}
          \item The cosine of the angle enclosed by the two daughters has to be physical, i.~e. in the region ($\{-1,1\})$. 
          \item The cosine of the angles between the two daughters and the mother must be in the same physical region. 
          \item In case, {\tt kinecheck()} was called from {\tt firstdice} not only the actual branch, but also all further branches 
                           have to be checked due to the possible changed kinematic of a propagator. This step is achieved using a 
                           recursion, where  {\tt kinecheck()} is called with the daughters as arguments.  
          \end{enumerate}
\item {\tt set\_q2\_z()} determines a new pair of virtuality and energy fraction $\{t,z\}$ for a daugther. First, the process 
          defining the daughters decay is specified with {\tt
n\_sudakov.set\_process()}. 
Then a new pair of $\{t,z\}$ is determined 
           with the help of {\tt n\_sudakov.av\_q2\_z()}. 

Now, this pair 
is subjected to a number of test.
           \begin{enumerate}
           \item If the virtuality has reached the cut-off mass $Q_0^2$ no further decay takes place.
           \item No new jet is allowed to emerge via parton--decays outside the matrix elements, see Subsec.\,\ref{physXSPS}.
                               The corresponding check is performed via {\tt jetfinder.jet\_con1()}. 
          \item Employing the virtuality ordered parton shower an angular ordering might be enforced ``by hand'' to 
                                account for the proper treatment of coherence effects. The actual opening angle is calculated and compared 
                               with the angle of the related predecessor.
         \item Taking mass effects into account a dead cone appears preventing collinear radiation of gluons off massive quarks. 
                               A cut in the phase space is enforced by definiting of a minimal opeing angle
         \end{enumerate}
         Having passed all the tests above the actual pair $\{t,z\}$ is accepted. Otherwise the calculation continues with the 
         determination of a new pair of daughters with the method {\tt n\_sudakov.av\_q2\_z()}.
\item {\tt set\_phi()} : The azimuthal angle between the plane of the two daughters and the plane spanned by the mother 
          and the granny is determined at this stage. Two options are at disposal, taking into account azimuthal correlations 
          or not. In the latter case the angle is distributed uniformly. The coefficient for the azimuthal correlations is calculated 
          with the help of the splitting functions. The method {\tt set\_ds()} is used to choose the appropriate
          {\tt dsec} with the outgoing flavour pair of the decay under consideration as input. Accordingly, {\tt dsec.differ()} 
          yields the value of $P(z)$.  The azimuthal angle is determined with a hit or miss method according to the evaluated 
          coefficient by virtue of the following algorithm.
          \begin{enumerate}
                   \item An angle $\phi$ is chosen uniformly in $2\pi$.
                   \item The value $f$ of the correlation factor is calculated with the angle $\phi$ entering. 
                   \item The procedure is repeated, until the ratio of $f/f_{\rm max}$ is bigger than a random number. 
          \end{enumerate}
\item {\tt set\_ds()} is used by {\tt set\_phi()}. It is responsible for choosing the correct splitting function according to the
          outgoing flavours.  
\end{enumerate}

\subsubsection{Sudakov form factors - {\tt n\_sudakov}}

\begin{figure}[h]
\centerline{\epsfxsize=14cm\epsffile{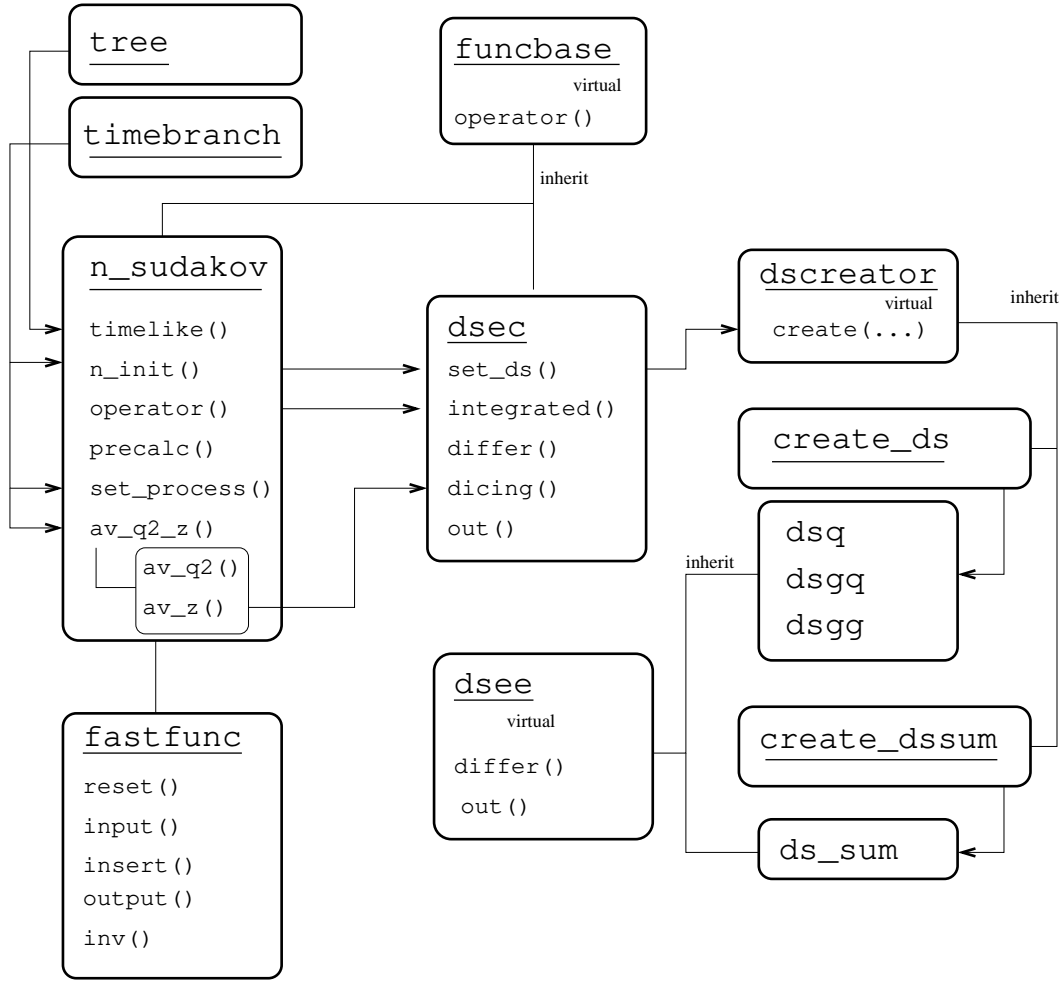}}
\caption{\label{nsudfig} The interplay between {\tt n\_sudakov} and the classes hosting the
splitting functions, organized via {\tt dsec}.}
\end{figure}

The class {\tt n\_sudakov} is responsible for the evaluation and use of the Sudakov form factors. Formally the class
is derived from the class {\tt func\_base}. The inheritance is due to the integration of the form factors employing an external 
routine. This method {\tt chebyshev()} integrates functions if they are represented by a {\tt func\_base}. {\tt n\_sudakov} is 
exclusively used by the class {\tt timebranch} and in turn utilizes the methods of {\tt dsec} for the evaluation of the splitting 
functions, see Fig.\,\ref{nsudfig}. The various methods for the pre-calculation and initialization read:
\begin{enumerate}
\item  For the calculation of the different Sudakov form factors the appropriate splitting functions are required. 
          In {\tt n\_init()} four different sums of them are created utilizing the class {\tt create\_dssum}. With {\tt dsec.set\_ds()}
          the splitting of a massless quark, a gluon, a charm quark and a beauty quark are initiated accordingly.
\item {\tt timelike()} maintains the pre-calculation of the different Sudakov form factors in the timelike region, i.~e. with 
         $q^2>q_0^2>0$ and is called from {\tt tree.init()}. 

         After the initialization of the appropriate single splitting functions with {\tt dsec.set\_ds()} {\tt precalc()} 
         determines a single form factor related to some specific decay $a\to bc$. The individual form factors 
         in turn are combined to the appropriate sum, where a Sudakov form factor for a massless quark without radiating
         photons, for massless up-- and down--type quarks including photon radiation, for gluons, charm quarks , and for beauty 
         quarks exist. Note, that in principle this method does not depend on the ordering scheme, which is adjusted at the 
         beginning of the calculation. 
\item {\tt precalc()} determines a look-up table of a Sudakov form factor, which depends on virtual masses in the scheme selected. 
         First, the method {\tt fastfunc.input()} is applied, checking for a possible read--in of the table. If the table
         is not available, the calculation starts. The method {\tt fastfunc.reset()} initiates the look-up table, which is derived from
         the class {\tt fastfunc}. A loop from the cut-off to the actual maximal virtuality, i.~e. the CMS energy, evaluates the
         form factor depending on the virtuality. With the {\tt chebyshev()} integration routine the appropriate values are calculated 
         and with {\tt fastfunc.insert()} they are stored in dependence on the actual scale. 
         The function {\tt chebyshev()} uses a pointer to the class {\tt n\_sudakov} for integration. This causes the reduction of this 
         class to its {\tt func\_base} part, whereas a {\tt func\_base} is a pure virtual class with only one method, {\tt operator()}. 
         Accordingly the function to be integrated is given by {\tt n\_sudakov.operator()}. At the end the table of Sudakov form factors 
         ist stored with {\tt fastfunc.output()}.        
\item {\tt operator()} yields the connection to the method {\tt dsec.integrated()}.  
\end{enumerate}

During event generation a number of methods is used for the determination of a new virtuality $t$, an energy fraction $z$ or the
outgoing flavour of a branch: 
\begin{enumerate}
\item {\tt set\_process()} is called from {\tt timebranch.set\_q2\_z()} and {\tt timebranch.dice()}. Depending on the ordering scheme and the 
          incoming flavour two pointers are set, one to the {\tt fastfunc} related to the corresponding Sudakov form factor and the second one 
          to the appropriate splitting function {\tt dsec}. Later they are responsible for the determination of the virtuality $t$ and the energy 
          fraction $z$, respectively. 
\item {\tt av\_q2\_z()} : At this stage a difference between the two ordering schemes is visible for the first time. For virtuality
          ordering the functions {\tt av\_q2()} and {\tt av\_z()} define the virtuality and the energy fraction, respectively. In angular 
          ordering {\tt av\_q2()} dices the energy scaled evolution variable $\tilde t$. The energy fraction is derived with {\tt av\_z()} as
          well, whereas the proper virtual mass of the decaying particle is calculated from $\tilde t$. A first kinematical check for the 
          energy fraction $z$ is enforced, where $z$ has to be in the range of $z_{\rm min}\le z \le 1-z_{\rm min}$ with 
          \be
          z_{\rm min} = \frac 12 \left(1-\sqrt{1-\frac t{w^2}}\,\right)\,.
          \ee
          Here $w$ represents the particle energy. If the value is not accepted the next pair $\{t(\tilde t),z\}$ is created starting from 
          the present value of $t(\tilde t)$.  
\item {\tt av\_q2()} determines a new virtuality with the actual Sudakov form factor. An inversion of the table with the method 
         {\tt fastfunc.inv()} yields the appropriate value of the virtuality.
\item {\tt av\_z()} is a link to the method {\tt dsec.dicing()}, which determines the energy fraction $z$. 
\item {\tt out()} yields the outgoing flavour pair of a branch utilizing the method {\tt dsec.out()}.
\end{enumerate}

\subsubsection{Splitting function - {\tt dsec}}

The splitting functions are organized within the class {\tt dsec}. The group of classes connected with {\tt dsec} shows a structure 
similar to the classes for the calculation of the cross sections, which are related to {\tt xsec}. Therefore the classes 
{\tt ds\_sum}, {\tt dscreator} and {\tt dsee} fulfill likewise tasks.  

{\tt dsec} is, in difference to {\tt xsec}, derivated from the virtual class {\tt func\_base}. As already explained in the context
of {\tt n\_sudakov}, this class is connected to the integration by the method {\tt chebyshev()}. {\tt dsec} is extensively used by the 
two classes {\tt timebranch} and {\tt n\_sudakov}. The appropriate methods are related to the integration of a splitting function 
or the dicing of an energy fraction $z$ obeying these splitting functions:
\begin{enumerate}
\item {\tt set\_ds()} : The method initializes a single or a list of splitting functions depending on the appropriate {\tt dscreator}. The
         method {\tt dscreator.create()} enables this task in analogy to its counterpart {\tt xscreator.create()}. 
         The incoming and outgoing flavours are used as arguments. Note, that in the case of a sum of splitting functions the 
         outgoing flavours are merely dummy variables and ignored accordingly.   
\item {\tt integrated()} yields the integrated splitting function including the factor of the strong coupling $\alpha_s[p_\perp^2(t,z)]$. 
         Utilizing the integration routine {\tt chebyshev()} and thereby the {\tt operator()}, the numerical integration is
         straightforward.  
\item {\tt operator()} calculates the value of the splitting function times the strong coupling $\alpha_s$ by means of the method 
           {\tt dsee.differ()}.
\item {\tt differ()} gives the value of the splitting function without the strong coupling constant. 
\item {\tt dicing()} is the central method for the determination of an energy fraction $z$ according to the splitting functions used. The
         value is diced with a hit or miss Monte Carlo method. Again, the {\tt operator()} is employed.
\item {\tt out()} returns the pair of outgoing flavours for a branch. The method {\tt dsee.out()} is utilized for this task.
\item {\tt set\_flav()} is used for setting the incoming flavour of a branch.
\item {\tt setang()} and {\tt setvirt()} are used for setting the angular or virtuality ordered parton shower. This effects the limits
         of the integration and the argument of the strong coupling constant. 
\end{enumerate}

The classes {\tt create\_ds} and {\tt create\_dssum} are derivated from the class {\tt dscreator}. They are responsible for choosing one
specific splitting function or for generating a list of splitting functions, respectively. The only method is {\tt create()}, where the input
parameters are the incoming and outgoing flavours. In the case of {\tt create\_dssum}, only the incoming flavour is
regarded. Hence, the list of splitting functions is build up for all processes with the same incoming flavour. They are combined with
respect to the switch {\tt sw.prompt\_gammas()} for including photon radiations and to the parameter {\tt pa.zflav()} for the 
maximum number of allowed flavours. The method {\tt create()} of the two classes is called from {\tt dsec.set\_ds()}.

The class {\tt ds\_sum} is responsible for the handling of a sum of splitting functions. It is derivated from the class {\tt dsee} and
contains a list of pointers to the appropriate splitting functions. Three methods are employed:
\begin{enumerate}
\item {\tt add()} allows for the addition of new splitting functions.
\item {\tt differ()} calculates the sum of splitting functions with {\tt dsee.differ()}.
\item {\tt out()} determines the two outgoing flavours of a branch. Therefore it first decides, which splitting function should be
          taken. The choice is made according to the value of the different $P(z)$ at the actual energy fraction $z$. Then the method 
          {\tt dsee.out()} returns the appropriate pair of flavour. 
\end{enumerate}

The different classes, which contain the splitting functions, are derivated from a virtual mother class {\tt dsee}. It defines the
standard methods for every splitting function class in analogy to the class {\tt xsee} encountered when discussing the
interfaces to the various matrix element generators. However, it contains only two methods. {\tt differ()} returns the value of the 
appropriate splitting function $P(z)$ at an energy fraction $z$. The pair of outgoing flavours is determined with the method {\tt out()}.  

Three different classes for the calculation of splitting functions exist, namely {\tt dsq}, {\tt dsgq}, and  {\tt dsgg} for the
splittings $q\to qg$, $g\to q\bar q$ and $g\to gg$, respectively. Derivated from the class {\tt dsee} the appropriate methods are
implemented accordingly. Special methods for regarding mass effects and radiation of photons are:
\begin{enumerate}
\item {\tt dsq} : The splitting function for radiating a photon from a quark differs from the appopriate radiation of a gluon by a 
         pre--factor. The methods {\tt QCDpref()} and {\tt QEDpref()} calculate the QCD ($\alpha_s\,C_F$) and the QED 
         ($\alpha_{\rm QED}\,e_q$) factor, respectively.  The branchings $q\to qg$ and $q\to gq$ can be derived by
         means of the methods {\tt diff\_qg()} and {\tt diff\_gq()}. Mass effects are included through cuts in the phase space available
         for the decay. This results in a change of the minimal and maximal value of the energy fraction $z$. The method is 
         implemented into the two routines {\tt diff\_qg\_m()} and {\tt diff\_gq\_m()} according to 
         the appropriate branching.  
\item {\tt dsgq} : The branching of a gluon into a massive quark-pair is accompanied by a proper cut in phase space for their 
         radiation preventing a gluon of, say, virtual mass $2$ GeV splitting into two $b$--quarks. The method {\tt diff\_m()}, in contrast 
         to {\tt diff()}, takes account of these effects.   
\item {\tt dsgg} : No additional method is needed.
\end{enumerate}

\subsection{Fragmentation\label{progHad}}

\subsubsection{Organizing with {\tt hadron}}

\begin{figure}[h]
\centerline{\epsfxsize=14cm\epsffile{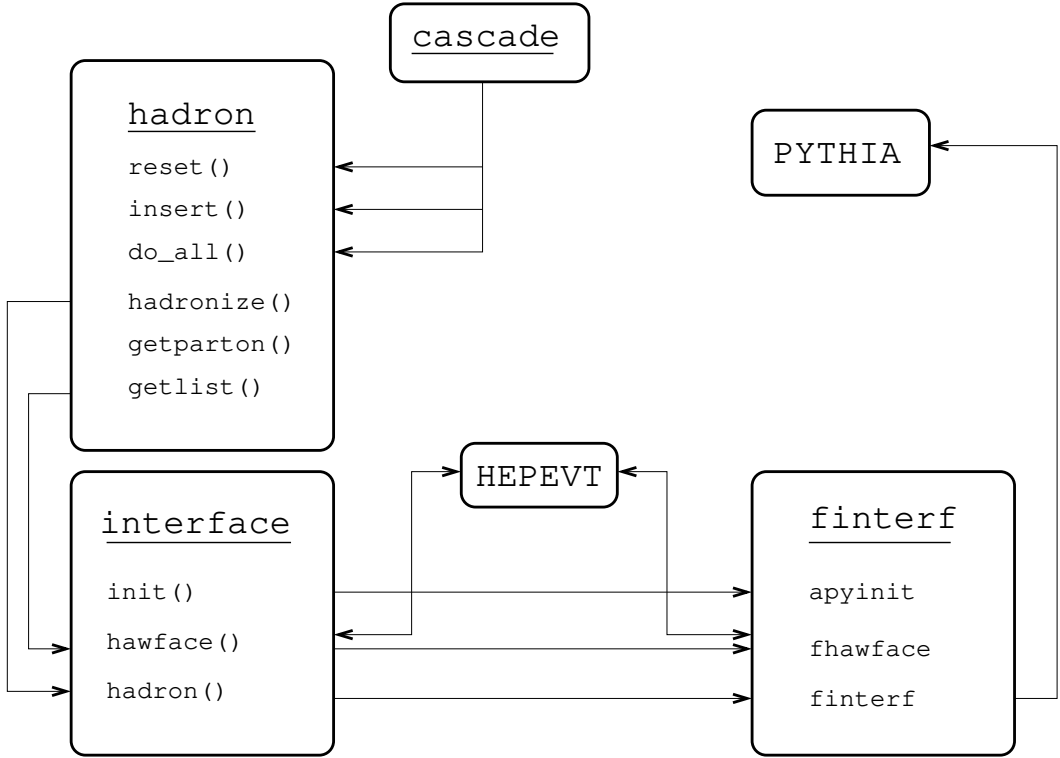}}
\caption{\label{hadfig} The interplay between {\tt hadron} and the {\tt interface} hosting methods to communicate with the 
{\tt Fortran}--interface routines and {\tt Pythia}.}
\end{figure}

The fragmentation of the partons is maintained by the class {\tt hadron}. In other words, within \APA this class currently organizes 
the connection to the fragmentation scheme of {\tt Pythia}, i.~e. the Lund string. For a scheme depicting the primary communication lines
between the classes and {\tt Fortran} routines see Fig.\,\ref{hadfig}. Since the process of fragmentation is carried out
there, the primary task is the construction of a list of partons, which are in a correct colour order. This is due to the fact, that a
string is always drawn between a quark and an anti-quark forming a colour singlet. Additional gluons are treated as kinks. 
Consequently, a quark cannot be connected to two strings, therefore care has to be taken, especially when a secondary $q\bar q$ pair
led to the break--up of the primary string into two. Then no gluons are allowed filling in the gap between the two strings, defining
some specified ordering of emitted gluons into the list. However, this list has to be filled into a {\tt C++} structure, which resembles 
the standardized {\tt HEPEVT} common block, see Tab.\,\ref{HEPEVT}. 

\begin{table}[h]
\begin{center}
\begin{tabular}{|l|l|}
\hline
NEVHEP    & number of events \\
NHEP      & number of entries\\
ISTHEP(I) & status of the entry\\
IDHEP(I)  & particle code \\
JMOHEP(I) & pointer to the first mother \\
JDAHEP(I) & pointer to the second mother \\
PHEP(5,I) & $(p_x,p_y,p_z,E,M)$ of this entry \\
VHEP(4,I) & $(x,y,z,t)$ from the production vertex of this entry\\
\hline
\end{tabular}
\caption{\label{HEPEVT} The structure of the {\tt HEPEVT} common block.}
\end{center}
\end{table}

The methods of {\tt hadron} are

\begin{enumerate}
\item {\tt reset()} : The list of partons is initialized with partons represented by their parton number.    
\item {\tt insert()} fills a pair of partons into the list, where the correct colour order has to be respected. A branching of a mother parton into two
          daughter partons is performed in the method {\tt cascade.branch()}. After the branch the parton number of the mother will be inherited to
          one of the daughters, the other one receives a new number. Then, the mother parton will be replaced by her daughters. The sequence of this 
          insertion into the list depend on the flavours of the daughters for the reasons named above:
          \begin{enumerate}
          \item $g\to gg$ : The two outgoing gluons can not be
distinguished anyhow, therefore no specified sequence is necessary.
          \item $q\to qg$ : The two cases of a quark or an anti-quark splitting are regarded differently, where the sequence of the insertion is 
                   $q -g$ or $g - \bar q$, respectively. The same holds true for the branching $q\to q V$, where $V$ stands for a photon, $Z$-, or
                   $W^\pm$-boson.  
          \item $g\to q\bar q$ : The insertion of a new quark pair is the most delicate task. This is due to the fact, that the sequence has to match with the
                    rest of the list, i.~e. strings can only be drawn between a quark and an anti-quark. Therefore first of all, the list has to be searched
                    upwards for the last quark insertion. In case it was a quark an anti-quark has to come first in the row and vice versa. Now, the two
                    partons can be included accordingly.    
          \end{enumerate}
         {\tt insert()} is called from {\tt cascade.epem\_init()} for the insertion of the first two partons and from {\tt cascade.branch()} for
         the further filling of the list.
\item {\tt do\_all()} is called from {\tt cascade.hadronize()} and maintains the fragmentation via a translation of the partons from the
           list of parton numbers into the appropriate {\tt HEPEVT} structure by {\tt hadronize()}. Now, the hadronization is performed in the external
           {\tt Fortran} code. The new list, which is now extended by the hadrons, is obtained via calling {\tt getlist()}.   
\item {\tt hadronize()} : The input of this method is the {\tt partlist} received from {\tt do\_all()}. Thereby, the method {\tt getparton()} yields the actual 
           parton in the list, which is now handed over to the {\tt C++} structure of the {\tt HEPEVT} common block. The flavour is transformed with the 
           method {\tt flavour.hepevt()} into the standardized {\tt HEPEVT} coding conventions. Having at hand the complete list of partons the 
           {\tt HEPEVT} structure is filled into its {\tt Fortran} counterpart with the method {\tt interface.hadron()}. This method performes the 
           fragmentation as well. 
\item {\tt getparton()} yields the parton for a given parton number.
\item {\tt getlist()} : This method returns the {\tt partlist}, which has been obtained from the {\tt HEPEVT} common block with {\tt interface.hawface()}. 
          Usually this method is called after the fragmentation.
\end{enumerate}

\subsubsection{Interface, {\tt C++} part\label{InterC}}

The class {\tt interface} provides methods for the fragmentation of the partons via and the run of the event generators available, i.~e. {\tt Pythia} 
and {\tt Herwig}. The {\tt Fortran} routines of the file {\tt finterf.f} establish the second part of the connection to the generators. Consequently,
the methods of {\tt interface} are: 

\begin{enumerate}
\item {\tt init()} is called from {\tt apacic.initapacic()} and initializes the fragmentation. For this task the {\tt Fortran} routine
          {\tt APYINIT} is called with the three important parameters of the Lund string fragmentation $a$, $b$, and $\sigma_q$ to be set by the user
          via {\tt parameter.dat}, for more details, see Sec. \ref{physHad}.    
\item {\tt hadron()} is used by the method {\tt hadron.hadronize()} and fills the list of partons from the {\tt C++} structure of the {\tt HEPEVT} common 
           block into its {\tt Fortran} counterpart via the subroutine {\tt FINTERF}.  
\item {\tt hawface()} obtains the contents of the {\tt HEPEVT} common block with the {\tt Fortran} routine {\tt FHAWFACE}. The structure of
           this common block is translated into a list of partons, which is appended to the incoming {\tt partlist}. The flavour ID's are
           transformed with {\tt flavour.from\_hadron()} and {\tt flavour.from\_hepevt()} for hadrons and partons respectively. Assigning the momentum 
           to the appropriate particles concludes this step. 
\item {\tt jetset()} generates one full event with {\tt Pythia} by means of the {\tt Fortran} routine {\tt APYRUN}. It is called by 
          {\tt apacic.jetset()}. 
\item {The same holds true for {\tt herwig()}, with {\tt Herwig} replacing {\tt Pythia}. The {\tt Fortran} routine {\tt AHWRUN} performes this task. }
\end{enumerate}

\subsubsection{Interface, {\tt Fortran} part}

The {\tt Fortran} part of the interface is dedicated to fill in and read--out the {\tt HEPEVT} common block. In addition to the routines
which are used for the fragmentation, others are provided for running the event generators {\tt Pythia} and {\tt Herwig}. 
The {\tt subroutines} cover the following tasks:
\begin{enumerate}
\item {\tt FINTERF} : The {\tt C++} structure of the {\tt HEPEVT} common block is filled into its {\tt Fortran} counterpart. Then, this common block is translated 
          into the internal {\tt Pythia} common block structure with the subroutine {\tt PYHEPC}. Now, {\tt PYEXEC} performs the fragmentation. The translation 
          back into the {\tt HEPEVT}--block is conducted with {\tt PYHEPC}, too. {\tt FINTERF} is exclusively  called by {\tt interface.hadron()}.   
\item {\tt FHAWFACE} reads-out the {\tt HEPEVT} common block. Thereby, the type of the different hadrons is translated into an {\tt APACIC++}
          internal pseudo-code, where charged or neutral mesons, baryons and leptons are encoded with the same number respectively. Elementary
          particles have their proper code in {\tt APACIC++} and remain therefore the same. The routine is called from {\tt interface.hawface()}. 
\item {\tt APYINIT} sets all initial parameters and switches for the fragmentation with {\tt Pythia}. It is called by {\tt interface.init()}. In the case of a 
          complete {\tt Pythia} run the initialization is executed with the subroutine {\tt PYINIT}. Note, that every switch and parameter deviating from the 
          standard Lund string parameters $a$, $b$ and $\sigma_q$ is not adjusted from {\tt APACIC++}. Therefore, this is the place for any 
          additional changes.     
\item {\tt APYRUN} : One complete event with {\tt Pythia} is performed in this routine by means of the subroutine {\tt PYEEVT}. After the run
           the routine {\tt PYHEPC} cares for the proper translation of the internal {\tt Pythia} common blocks into {\tt HEPEVT}. {\tt APYRUN} is 
           used by {\tt interface.jetset()}.
\item {\tt AHWINIT} initializes a {\tt Herwig} run. First, the different switches and parameters, like the process number {\tt IPROC}
           or the beam energies have to be set. Note, that any adjustments have to be made in this routine. Then the subsequent calls to the
           methods {\tt HWIGIN}, {\tt HWUINC}, and {\tt HWEINI} perform the initialization.  
\item {\tt AHWRUN} performes one event with the event generator {\tt Herwig}. It is called by the method {\tt interface.herwig()}.
\end{enumerate}

\newpage
\section{Installation guide\label{inst}}

\subsection{Installation}

Here we want to explain briefly how to install \APA. Since \APA has only a small selection of included matrix elements, the user might want to
use it together with \AME, our preferred matrix element generator. Therefore we describe the installation of the complete package with emphasis on 
the \APA part. Further details on \AME can be found in \cite{AMEGIC}.

At the moment, the package \AA can be obtained upon request from the authors, a homepage for downloads is in preparation. After unpacking with 
the usual {\tt tar} command the two directories {\tt APACIC++-1.0} and {\tt AMEGIC++-1.0} as well as a script {\tt install} will be generated. 
This script includes the call to another script for the automatic configuration of the Makefiles and their execution. In the best of all worlds, a simple 
call of {\tt install} is sufficient. Then the executable {\tt apacic} is ready to use in the directory {\tt APACIC++-1.0/apacic}.  

If this script does not work, an adjustment of the Makefiles has to be made by hand. In case the generated Makefiles could not be used, the
alternative script {\tt pseudomake} is at disposal. It compiles and links the program files hard-wired. Therefore it has to be edited for
setting the compilers and their options. The script is preconfigured for the GNU compiler family. 

During compilation and linking a second problem can occur, due to the slightly different approach to standard classes of different
compilers. These classes are especially the two build-in classes {\tt complex} and {\tt string}. The first troubles might appear, if the class
name differs from the name used in the program. A simple redefinition with {\tt typedef} solves this problem. Since, these changes should not be 
made in every program part relying on these classes two pseudo header files are used instead of the standard headers. They are
located in the {\tt include} directory. All necessary adjustments can be made in these header files, named {\tt mycomplex} and {\tt mystring}.  
The class {\tt complex} could often be fitted by simply renaming the class, whereas the {\tt string} class can also differ in the
definition of the different methods. Therefore we programmed an own {\tt string} class, which can be used, if the standard class fails. 

The package \AA\ uses different matrix element generators and hadronization models, which are linked to the program. Most of them are written 
in {\tt Fortran}. Accordingly the {\tt Fortran} standard libraries have to be used, which may also differ on different machines or compilers. If needed, these 
changes should be made in the {\tt Makefile}, which can be found in the directory {\tt APACIC++-1.0/apacic}.    

Any successful installation procedure results in the executable {\tt apacic} to be found in the subdirectory {\tt APACIC++-1.0/apacic}.

\subsection{Running \APA\label{Run}}

The physics encoded in \APA is controlled by a number of parameters and switches. They are read in during the program execution and can be
found in the two files {\tt parameter.dat} and {\tt particle.dat}, which have ASCII format and can therefore be edited easily. Both {\tt dat} files
can be found in the subdirectory {\tt APACIC++-1.0/apacic}. The first one is connected to the parameters and switches, whereas the second
file contains all particle related data. A sample and some explanations to the different parameters and switches can be found in
the Tables \ref{param} and \ref{switch}. The procedure of how to declare particle masses, widths etc. is exemplified in Table \ref{particle}. Note,
that the steering of the matrix element generator \AME is already included, but the use of some parameters might need additional changes, see
\cite{AMEGIC}.

\begin{table*}[p]
\begin{center}
\caption{Parameters in \APA}
\label{param} 
\begin{tabular}{llll}
\hline
type & name & default & meaning\\
\hline
int & jobnumber & 100 & number of the job\\
double & ws & 91.1884 & center of mass energy in GeV\\
int & nev & 10000 & number of events\\
double & err & 0.01 & Error by calculating matrix-elements\\
int & jet & 3 & maximal number of jets \\
&&& -- exclusive $n$--jet production is selected by $10\cdot n$\\
int & zflav & 5 & number of flavours\\
\hline
double & asMZ & 0.118 & $\alpha_S$ on the Z--pol\\
double & kappaS3 & 0.01 & $\kappa_S^3$ for $3$ jets \\
double & kappaS4 & 0.01 & $\kappa_S^4$ for $4$ jets\\
double & kappaS5 & 0.01 & $\kappa_S^5$ for $5$ jets\\
double & q02 & 0.2 & parton shower cut--off\\
double & Lqcd & 0.16 & $\Lambda_{\rm QCD}^2$\\
double & aqed & $1/128$ & $\alpha_{\rm QED}$\\
double & SW & $\sim\sqrt{0.22}$ & $\sin{\theta}$ -- the Weinberg angle, in the
parameter file $\sin^2{\theta}$   \\
double & ycut\_ini & 0.01 & $y_{\rm cut}$ for jet clustering in the
initial state\\
double & ycut\_fin & 0.01 & $y_{\rm cut}$ for jet clustering in the
final state\\
double & Lund\_a & 0.358 & parameter $a$ for Lund string
hadronization\\
double & Lund\_b & 0.850 & parameter $b$ for Lund string
hadronization\\
double & Lund\_sigma & 0.372 & parameter $\sigma_q$ for Lund string
hadronization\\ 
\hline
\end{tabular}
\end{center}
\end{table*}

\begin{table*}[p]
\begin{center}
\caption{Switches in \APA\ part I}
\label{switch} 
\begin{tabular}{lcl}
\hline
name & defaults & meaning\\
\hline
generator & 1 & use \APA=1, {\tt JETSET/PYTHIA}=2 or\\
&& {\tt HERWIG}=3\\
amegic & 0 & use \AME\ interface on=1,off=0\\
debrecen & 0 & use {\tt DEBRECEN} interface off=0, LO=1 and NLO=2\\
excalibur & 0 & use {\tt EXCALIBUR} interface on=1,off=0\\
QCD & 1 & pure QCD jet production on=1,off=0\\
ew & 0 & electroweak four jets on=1,off=0\\  
(gam,z,w,h)decay & 1 & $\gamma,Z,W^\pm,H$ decay on in the matrix elements\\
isr    & 0 & initial state radiation\\
massiveME& 0 & massive quarks in the matrix element on=1,off=0\\
massivePS& 0 & massive quarks in the parton shower on=1,off=0\\
massrun& 0 & running masses on=1,off=0, see \cite{MassWidths}\\
width  & 0 & running width  off=0,, Exc.=1,s-dep=2\\
runQED & 0 & running $\alpha_{\rm QED}$ on=1,off=0\\
coulomb& 0 & Coulomb corrections on=1,off=0, see \cite{WWCoulomb}\\
multichannel & 0 & use Rambo=0/Multichannel=1 for\\
&& phasespace generation\\
jetinitial & 1 &  {\tt DURHAM}=1, {\tt JADE}=2 or {\tt GENEVA}=3 as\\
&& initial jet clustering scheme\\
rescale & 1 & direct=0/rescaled(1)=1/rescaled(2)=2/\\
&&NLL-matched=3 jet rates\\
probabs & 0 & way to evaluate probabilities\\
&&  without=0,with=1 interferences\\
\hline
\end{tabular}
\end{center}
\end{table*}
\begin{table*}[p]
\begin{center}
\caption{Switches in \APA\ part II}
\begin{tabular}{lcl}
\hline
name & defaults & meaning\\
\hline
shower & 2 & parton shower off=0, angular ordering=1\\
&& or order by virtualities=2\\
a\_crit & 1 & angular ordering within virtuality \\
&& ordered parton shower on=1,off=0\\
azim & 1 & azimuthal correlations within \\
&& the parton shower on=1,off=0\\
prompt\_gammas & 1 & include prompt photons within shower on=1,off=0\\
hadron & 1 & hadronization off=0, by {\tt JETSET/PYTHIA}=1\\
analysis & 1 & event analysis on=1,off=0\\
jetfinal & 1 &  {\tt DURHAM}=1,{\tt JADE}=2 or {\tt GENEVA}=3 as \\
&& final jet clustering scheme\\
kappaS\_var & 2 & $\kappa_S$ scaling no=0,$4$ jet=1,$3+4+5$ jet=2, $4+5$
jet=3\\   
kin\_match & 2 & combining the kinematics direct=0/$\alpha_S$=1/NLL=2\\
\hline
\end{tabular}
\end{center}
\end{table*}

\begin{table*}[p]
\caption{Particle in \APA}
\label{particle} 
\begin{tabular}{rrrrrr}
\hline
kf-code & Mass & Width & 3*Charge & Weak isospin & Name\\
1 & .01	    & .0    & -1 & -1 & d\_quark\\
2 & .005    & .0    & 2	 & 1  & u\_quark\\
3 & .2	    & .0    & -1 & -1 & s\_quark\\
4 & 1.7     & .0    & 2	 & 1  & c\_quark\\
5 & 4.7     & .0    & -1 & -1 & b\_quark\\
6 & 175.0   & .0    & 2	 & 1  & t\_quark\\
21& .0	    & .0    & 0	 & 0  & gluon\\
22& .0	    & .0    & 0	 & 0  & photon\\
23& 80.356  & 2.07  & -3 & 0  & W-\\
24& 91.188  & 2.439 & 0	 & 0  & Z\\
25& 120.0   & 10.0  & 0  & 0  & Higgs\\    
31& .000512 & .0    & -3 & -1 & e-\\
32& .0      & .0    & 0  & 1  & nu\_e\\
33& .105    & .0    & -3 & -1 & mu-\\
34& .0      & .0    & 0  &  1 & nu\_mu\\
35&1.776    & .0    & -3 & -1 & tau-\\
36& .0      & .0    & 0  &  1 & nu\_tau\\
\hline
\end{tabular}
\end{table*}

After the specification of the parameters the next question to be tackled is how to start the program. Different ways for
three programming languages, will be presented. 

Of course, the easiest way to run \APA is to start it under {\tt C++}. A sample main program, {\tt apacic.C}, is already included in
the package. It makes use of the general object oriented structure:\\ 

{\tt 
//  apacic.C: main() function\\
\\
\#include "vec.H"\\
\#include "apacic.H"\\
\#include "param.H"\\
\\
extern parameter pa;\\
\\
int main() \{  \\
\vp$\qquad$ apacic ee;\\
\vp$\qquad$ ee.init();\\
\vp$\qquad$ ee.set\_s(pa.ws()*pa.ws());\\
\vp$\qquad$ ee.set\_count(pa.nev());\\
\vp$\qquad$ switch(sw.generator()) \{\\
\vp$\qquad\qquad$ case 1:  ee.Elektron\_Positron();break;\\
\vp$\qquad\qquad$ case 2:  ee.jetset();break;\\
\vp$\qquad\qquad$ case 3:  ee.herwig();break;\\
\vp$\qquad\qquad$ default: ee.Elektron\_Positron();\\
\vp$\qquad$ \}\\
\}\\
}

For running \APA from plain {\tt C} or {\tt Fortran} an interface is provided with three functions. One for initializing {\it apainit}, one
for handling one event {\it aparun} and one for finishing {\it apaend}. Thus, the typical {\tt C} file would look as follows\\

{\tt
int main() \{  \\
\vp$\qquad$ apainit();\\
\vp$\qquad$ long int NEVENT = 100000;\\
\vp$\qquad$ short int i;\\
\vp$\qquad$ for (i=0;i<NEVENT;i++) aparun();\\
\vp$\qquad$ apaend();\\
\vp$\qquad$ \}\\
\}\\
}

and the same for {\tt Fortran}\\

{\tt 
program apacic\\
\vp$\qquad$ call apainit\\
\vp$\qquad$ NEVENT = 100000\\
\vp$\qquad$ do i=1,NEVENT\\
\vp$\qquad\qquad$ call aparun\\
\vp$\qquad$ enddo\\
\vp$\qquad$ call apaend\\
END
}

The results obtained from \APA are availabe in various ways. First of all the calculated matrix elements are stored in the directory 
{\it me}. The included analysis tools, if switched on in {\tt parameter.dat}, are capable of generating histograms for different
observables, like event shapes. Eventually, the corresponding files can be found in the directory {\tt output}. Note, that in case the
hadronization of {\tt Pythia} is used, the internal common blocks like {\tt PYDATn} are already filled. Therefore external analysis tools,
which depend on these structures, can be used directly. The relevant informations can be obtained in every step after calling {\tt aparun}.  

\clearpage   
\newpage

\section{Summary\label{Summ}}

In this paper we have presented in some detail the new event generator \APA dedicated to the simulation of full $e^+e^-$ annihilation 
events at LEP and beyond. One of the cornerstones of \APA is the newly developped algorithm to combine arbitrary matrix elements
with the subsequent parton shower thus allowing for a good description of multijet events in a broad energy range. The main
difference to other event generators is, that channels with varying numbers of jets can be treated simultanously by means of
the corresponding matrix elements describing their production combined with the parton shower modelling their evolution. 
This has not been accomplished before and opens new perspectives for event generators for jet physics. In addition to some built--in 
matrix elements and the newly programmed matrix element generator \AME \APA provides interfaces to two further codes, allowing 
for the description of a large number of different channels and mutual checks of these programs. Of course, additional matrix elements 
can be linked, too, if the corresponding interfaces are provided.

For the parton shower modelling the subsequent evolution of the jets down to the regime of fragmentation via multiple
emission of secondary partons, two different schemes are made available within \APA, namely the LLA scheme or ordering by virtualities
and the MLLA scheme or ordering by angles. In the first scheme, the effect of coherence on the radiation pattern can be 
incorporated by vetoing on rising opening angles of subsequent branchings. These veto can be switched on and off as well
as the azimuthal correlations connecting the decay planes of subsequent splittings. Thus, \APA contains both schemes
for the parton shower in a state of the art--fashion. Both parts, i.~e. matrix elements and parton shower allow for the inclusion 
of mass effects, provided the matrix element generator includes them.

The fragmentation is currently carried out by means of the Lund--string. For this purpose, \APA uses a well--established other 
program, namely {\tt Pythia}, via a corresponding interface. In principle, other fragmentation schemes, like the cluster scheme
of {\tt Herwig}, could be used as well. The interface is currently under construction.

In \APA, initial state radiation, parton shower and fragmentation can all be switched on and off separately, for the hard processes
currently up to five jets can be treated, either inclusively or exclusively. The built--in tools for event analysis monitor
the final states as provided by the matrix element generators, the partons after the subsequent jet--evolution and finally
the hadrons after fragmentation and further hadronic decays. This allows for easy access to the individual stages of event generation
and corresponding checks.

As it stands, \APA has been developped from schratch in the modern computer language {\tt C++}. It incorporates roughly 10000
lines of code in {\tt C}-- and header--files, which are organized in more than 70 classes. It provides interfaces to another {\tt C++}--code
and two {\tt Fortran}--programs and has been tested successfully on a variety of plattforms, {\tt AIX}, {\tt Digital Unix} and {\tt IRIX}.
It has been developped on a Pentim II running under {\tt Linux}. Avoiding the {\tt template}--structures within \APA a rather high 
transportability of the code has been achieved. 

Within {\tt APACIC++}, we have made use of the object--oriented features of {\tt C++} which allow for the transparent organization of the code 
in classes and a good control of the data flow between them and their methods. The fact that large parts of typical event structures
can be translated into classes leads to an quick first understanding of the code and enables an easy access for the user. 
Additionally, the intensive use of inheritance defines standard methods for similar classes with identical purpose and
introduces further structure into the program. Recursions encountered in various places allow for the reduction to the basic
building blocks of algorithms and add their piece to a comprehensible programming style. Taken together we want to express 
our hope, that the abstract programming style made possible by using {\tt C++} leaves potential users in the position to
understand in some detail the algortihms encoded in \APA.

Within the framework of $e^+e^-$ events, some open tasks within \APA to be tackled in the future include

\begin{enumerate}
\item a better treatment of initial state radiation of photons off the electrons,
\item the inclusion of $\gamma\gamma$ and $\gamma e$ events, and 
\item an interface to the cluster fragmentation provided by {\tt
Herwig} \cite{CLUSTER}.
\end{enumerate}

In addition, we feel that there is an urgent need to go beyond $e^+e^-$ collisions and turn to $pp$ processes. 

\section*{Acknowledgements}

R.K. and F.K.  would like to thank J. Drees, K. Hamacher and U. Flagmeyer for helpful discussions. During the process of 
tuning the \APA--parameters to experimental data by U. Flagmeyer, we were able to identify and cure some shortcomings and 
bugs of the program. 

For R.K. and F.K. it is a pleasure to thank L. Lonnblad and T. Sjostrand for valuable comments and S. Catani and 
B. Webber for the pleasant collaboration on the combination of matrix elements and parton showers.

R.K. acknowledges the kind hospitality of the Technion, where parts of this work have been finalized.

This work is supported by BMBF, DFG, GSI and Minerva.

\newpage
\section*{Test run output}
Finally, we want to give an examplatory {\it Test Run Output} of \APA. Using the parameters, switches and particle data given above,
Sec.\,\ref{inst}, the output of \APA should look like this:  

{\it
Initializing APACIC++ 3-Jets (d\_quark,gluon,anti-d\_quark)\\
Integration Channels: \\
1 RAMBO: d\_quark;gluon;anti-d\_quark;\\
Relevant ID in look-up table is :3PZG\_2qd\_g\\
No file, no cross section ... \\
APACIC : Starting the calculation. Lean back and enjoy ... .\\
Integration Channels: \\
1 RAMBO: d\_quark;gluon;anti-d\_quark;\\
5000. LO-3-Jet: 1739.09 pb +- 2.42268\%\\
10000. LO-3-Jet: 1767.4 pb +- 1.69096\%\\
15000. LO-3-Jet: 1794.57 pb +- 1.36723\%\\
20000. LO-3-Jet: 1770.8 pb +- 1.17895\%\\
25000. LO-3-Jet: 1776.28 pb +- 1.05156\%\\
30000. LO-3-Jet: 1778.54 pb +- 0.96236\%\\
Cross section = 1778.54 pb +- 0.0096236\%, max = 1.15995\\
Initializing APACIC++ 3-Jets (u\_quark,gluon,anti-u\_quark)\\
Integration Channels: \\
1 RAMBO: u\_quark;gluon;anti-u\_quark;\\
Relevant ID in look-up table is :3PZG\_2qu\_g\\
No cross section 3PZG\_2qu\_g found ... \\
APACIC : Starting the calculation. Lean back and enjoy ... .\\
Integration Channels: \\
1 RAMBO: u\_quark;gluon;anti-u\_quark;\\
5000. LO-3-Jet: 1395.63 pb +- 2.35564\%\\
10000. LO-3-Jet: 1410.91 pb +- 1.64697\%\\
15000. LO-3-Jet: 1415.78 pb +- 1.3517\%\\
20000. LO-3-Jet: 1420.02 pb +- 1.17471\%\\
25000. LO-3-Jet: 1422.37 pb +- 1.04681\%\\
30000. LO-3-Jet: 1421.41 pb +- 0.957339\%\\
Cross section = 1421.41 pb +- 0.00957339\%, max = 0.9133\\
Initializing APACIC++ 3-Jets (s\_quark,gluon,anti-s\_quark)\\
Integration Channels: \\
1 RAMBO: s\_quark;gluon;anti-s\_quark;\\
Relevant ID in look-up table is :3PZG\_2qd\_g\\
Found the cross section in look-up file ./me/me\_91\_10\_D\_APACIC\\
Cross section = 1778.54 pb +- 0.0096236\%, max = 1.15995\\
Initializing APACIC++ 3-Jets (c\_quark,gluon,anti-c\_quark)\\
Integration Channels: \\
1 RAMBO: c\_quark;gluon;anti-c\_quark;\\
Relevant ID in look-up table is :3PZG\_2qu\_g\\
Found the cross section in look-up file ./me/me\_91\_10\_D\_APACIC\\
Cross section = 1421.41 pb +- 0.00957339\%, max = 0.9133\\
Initializing APACIC++ 3-Jets (b\_quark,gluon,anti-b\_quark)\\
Integration Channels: \\
1 RAMBO: b\_quark;gluon;anti-b\_quark;\\
Relevant ID in look-up table is :3PZG\_2qd\_g\\
Found the cross section in look-up file ./me/me\_91\_10\_D\_APACIC\\
Cross section = 1778.54 pb +- 0.0096236\%, max = 1.15995\\
Initializing APACIC++ 2-Jets (d\_quark,anti-d\_quark)\\
Integration Channels: \\
1 RAMBO: d\_quark;anti-d\_quark;\\
Relevant ID in look-up table is :2PZG\_2qd\\
No cross section 2PZG\_2qd found ... \\
APACIC : Starting the calculation. Lean back and enjoy ... .\\
Integration Channels: \\
1 RAMBO: d\_quark;anti-d\_quark;\\
5000. LO-2-Jet: 9262.7 pb +- 0.397043\%\\
Cross section = 9262.7 pb +- 0.00397043\%, max = 0.709496\\
Initializing APACIC++ 2-Jets (u\_quark,anti-u\_quark)\\
Integration Channels: \\
1 RAMBO: u\_quark;anti-u\_quark;\\
Relevant ID in look-up table is :2PZG\_2qu\\
No cross section 2PZG\_2qu found ... \\
APACIC : Starting the calculation. Lean back and enjoy ... .\\
Integration Channels: \\
1 RAMBO: u\_quark;anti-u\_quark;\\
5000. LO-2-Jet: 7213.25 pb +- 0.365187\%\\
Cross section = 7213.25 pb +- 0.00365187\%, max = 0.529516\\
Initializing APACIC++ 2-Jets (s\_quark,anti-s\_quark)\\
Integration Channels: \\
1 RAMBO: s\_quark;anti-s\_quark;\\
Relevant ID in look-up table is :2PZG\_2qd\\
Found the cross section in look-up file ./me/me\_91\_10\_D\_APACIC\\
Cross section = 9262.7 pb +- 0.00397043\%, max = 0.709496\\
Initializing APACIC++ 2-Jets (c\_quark,anti-c\_quark)\\
Integration Channels: \\
1 RAMBO: c\_quark;anti-c\_quark;\\
Relevant ID in look-up table is :2PZG\_2qu\\
Found the cross section in look-up file ./me/me\_91\_10\_D\_APACIC\\
Cross section = 7213.25 pb +- 0.00365186\%, max = 0.529516\\
Initializing APACIC++ 2-Jets (b\_quark,anti-b\_quark)\\
Integration Channels: \\
1 RAMBO: b\_quark;anti-b\_quark;\\
Relevant ID in look-up table is :2PZG\_2qd\\
Found the cross section in look-up file ./me/me\_91\_10\_D\_APACIC\\
Cross section = 9262.7 pb +- 0.00397043\%, max = 0.709496\\
Electroweak Multiquarks(>=4): 0\\
QCD         Multijets       : 42214.6\\
Sigma\_all: 1.\\
Partial Rates : \\
2 jet rate: 0.806265\\
3 jet rate: 0.193735\\
1. Event\\
...\\
10000. Event
}

This test run output can appear in a modified form, if the Sudakov form factors are not precalculated and stored
in the corresponding files. 

\end{document}